\documentclass[%
 amsmath,amssymb,
 prfluids,
]{revtex4-2}

\usepackage{newtxtext}
\usepackage{newtxmath}
\usepackage{natbib}
\usepackage{graphicx}
\usepackage{amsmath}
\usepackage{amssymb}
\usepackage{soul}
\usepackage{caption}
\usepackage{natbib}
\usepackage{soul}
\usepackage{color}
\usepackage{hyperref}
\hypersetup{
    colorlinks = true,
    urlcolor   = blue,
    citecolor  = black,
}
\newcommand{\RomanNumeralCaps}[1]

\def\black #1{\textcolor{black}{#1}}

\definecolor{burgundy}{rgb}{0.5, 0.0, 0.13}

\definecolor{past}{rgb}{0.47,0.87,0.47} %Green
\def\past#1{\textcolor{past}{#1}}
\def\bea{\begin{equation}}
\def\eea{\end{equation}}

\newcommand{\pdiff}[2]{\frac{\partial #1}{\partial #2}}

\newcommand{\soltype}[1]{\textsc{\romannumeral #1}}

\begin{document}

\preprint{APS/123-QED}

\title{On the stability of fully nonlinear hydraulic-fall solutions to the forced water-wave problem}

% Force line breaks with \\
\author{J. S. Keeler}
\email{j.keeler@uea.ac.uk}
\affiliation{School of Mathematics, University of East Anglia, Norwich, NR4 7TJ, UK}%Lines break automati
% Force line breaks with \\
\author{M. G. Blyth}
\email{m.blyth@uea.ac.uk}
\affiliation{School of Mathematics, University of East Anglia, Norwich, NR4 7TJ, UK}%Lines break aumati

%\maketitle
\begin{abstract}
Two-dimensional free-surface flow over localised topography is examined with the emphasis on the stability of hydraulic-fall solutions. A Gaussian topography profile is assumed with a positive or negative amplitude modelling a bump or a dip, respectively. 
Steady hydraulic-fall solutions to the full incompressible, irrotational Euler equations are computed, and their linear and nonlinear stability is 
analysed by computing eigenspectra of the pertinent linearised operator and by solving an initial value problem. The computations are carried out 
numerically using a specially developed computational framework based on the finite element method. The Hamiltonian structure of the problem is demonstrated and stability is determined by computing 
eigenspectra of the pertinent linearised operator. It is found that a hydraulic-fall flow over a bump is spectrally stable. The 
corresponding flow over a dip is found to be linearly unstable. In the latter case, time-dependent simulations show that the flow ultimately settles into a time-periodic motion that corresponds to an invariant solution in an appropriately defined phase space. Physically, the solution consists of a 
localised large amplitude wave that pulsates above the dip while simultaneously emitting nonlinear cnoidal waves in the upstream direction and 
multi-harmonic linear waves in the downstream direction.
%Whilst steady hydraulic-fall solutions for a bump 
%are well-known, the corresponding flow over a dip has received much less attention. Moreover, the stability of these solutions (in particular for the dip) has not received little theoretical treatment and, it seems, has not been subject to experiment.  finite-element formulation, a detailed theoretical/numerical linear stability analysis and a numerical nonlinear stability analysis indicates that hydraulic-falls over topographic dips are unstable. However, by examining the long-time evolution we identify a new stable time-dependent invariant solution of the forced irrotational Euler equations, consisting of a time-dependent wave-pulse at the centre of the forcing that emits cnoidal waves far upstream and and multi-harmonic linear waves downstream. We identify a hierarchy of commensurate temporal frequencies, the dominant frequency linked to the underlying point spectrum of the steady-state, which indicates this new invariant solution is a periodic-orbit.\textbf{ \past{JSK: words:190/250}}
\end{abstract}
\maketitle
\section{Introduction}

\begin{figure}
  \centering
  \includegraphics[width=\textwidth]{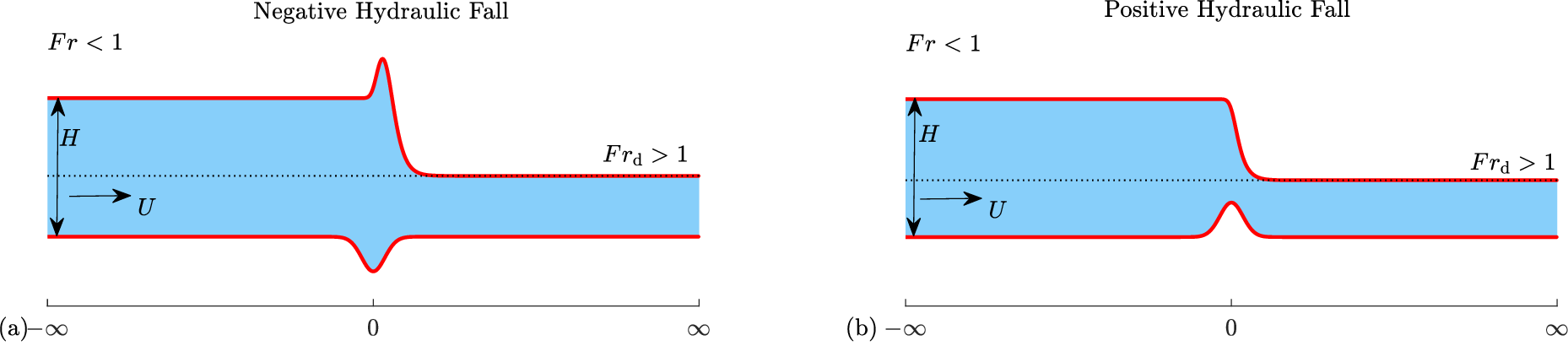}
  \caption{Hydraulic-fall solutions. Sketch of the two basic types in the non-dimensional domain, the arrow indicates the flow direction. (a) Bump forcing, $Fr<1$ (b) Dip forcing, $Fr<1$ where $Fr$ is the Froude number, defined in \eqref{froude}. The downstream Froude number, $Fr_{\mathrm{d}}$, is of opposite criticality to $Fr$. }
  \label{fig:summary}
\end{figure}

The incompressible flow of an inviscid liquid with a free surface over topography and/or subject to a surface pressure distribution is a 
classical problem in fluid dynamics that has received considerable attention over the last two hundred years \citep[see, for example][]
{whitham,akylas,dias+vanden-broeck}. The principal aim is to determine the free-surface profile and to explore how it changes as the relevant 
parameters, such as the oncoming flow speed or the amplitude of the forcing (either the topography or pressure distribution) are varied. For a localised forcing, it is well known that in one 
possible steady flow configuration, herein referred to as a hydraulic fall, the fluid level drops from a higher uniform level upstream to a lower uniform level downstream (figure~\ref{fig:summary}). Such flows have been observed experimentally \cite[e.g.][]{forbes1982,tam}, and have 
been computed as solutions to the fully nonlinear irrotational Euler (hereinafter FNL) equations \cite[e.g.][]{forbes1988critical,dias1989open}, and 
as solutions to reduced\past{-}order models such as the forced Korteweg-de Vries (hereinafter fKdV) equation \citep[see, for example,][]
{forbes1982,dias+vanden-broeck}.  A recent review is provided by \cite{binderreview}.

Most of the studies in the literature that concern hydraulic-fall solutions focus on steady flow, and there appears to have been very little effort 
devoted to gaining a theoretical understanding of the stability of these flows. Experimental studies have tended to concentrate on flow over a 
bump as depicted in figure~\ref{fig:summary}(b), and, indeed, we have been unable to identify any experimental work on hydraulic falls over a dip 
(figure~\ref{fig:summary}(a)). This may simply be due to the fact that presumably a wave tank with a bump is easier to engineer than one with a 
dip. In this article we discuss the stability properties of hydraulic-fall solutions for both positive and negative forcings\past{;} (i) by carrying out a linear stability analysis, and (ii) by developing a computational 
framework based on the finite-element method that we use to analyse nonlinear stability. We show that over a bump the 
hydraulic-fall solution is spectrally stable, but over a dip it is linearly unstable. Most intriguingly, from our suite of time-dependent 
calculations we identify a new time-dependent invariant solution of the FNL system which corresponds to a stable time-periodic orbit in a suitably 
defined phase space. 
%%%%%%%%%%%
%%%%%%%%%%%
    
The fKdV and FNL steady solution spaces for various topographic forcings are by now well established \cite[e.g.][]{forbes1982,
dias+vanden-broeck,wade+binder,wade+binder2}. For the fKdV system a number of different solution types can be identified 
\citep{akylas,binder+mccue,dias+vanden-broeck}; for a recent review see \cite{binderreview}. Broadly speaking, these can be classified as  cnoidal 
wave solutions which are flat upstream and wavy downstream, solitary-wave solutions which are flat both upstream and downstream, and 
transcritical solutions which we are herein referring to as hydraulic falls. These solutions are characterised by the upstream depth-based Froude number given by 
\begin{equation}
  Fr = \frac{U}{\sqrt{gH}},
  \label{froude}
\end{equation}
where $g$ is the gravitational constant, and $U$ and $H$ are the speed of the flow and the fluid depth far upstream. Cnoidal-wave solutions 
occur when $Fr<1$ (subcritical flow), and solitary-wave solutions occur when $Fr>1$ (supercritical flow). A distinguishing feature of hydraulic-fall 
solutions is that the flow upstream is subcritical and the flow downstream is supercritical. For a given topographic forcing, a steady hydraulic-fall solution exists only for a particular value of $Fr$. Furthermore, for hydraulic-fall solutions\past{,} if the forcing is negative (a dip) then the wave profile has a local maximum above the forcing, 
and if the forcing is positive (a bump) the fluid depth decreases monotonically from upstream to downstream (see 
figure~\ref{fig:summary}). 

Using numerical simulations of the fKdV system carried out for a number of different types of perturbations to the base steady state,
\cite{donahue}, \cite{chardard} and \cite{choi2016hyperbolic} all concluded that a hydraulic fall over a positive forcing is stable. 
\cite{page2014} showed via time-dependent calculations starting from a small wavy Gaussian perturbation, that the hydraulic fall solution appeared to be stable although they did not track the long-term behaviour. None of these studies attempted a formal linear stability analysis.

The article proceeds as follows. In 
\S~\ref{sec:problem} we state the mathematical problem and the weak formulation of the problem that forms the basis of our numerical 
framework. In \S~\ref{sec:fnl} we briefly describe the steady bifurcation structure of the FNL system. We carry out a linear stability analysis in 
\S~\ref{sec:linear_stability} and a nonlinear stability analysis via direct numerical simulations of the FNL 
model in \S~\ref{sec:nonlinear_stability}. Finally in \S~\ref{sec:unsteady_inverse} and \S~\ref{sec:discussion} 
we discuss our results and consider possible avenues for future research.

\section{Problem formulation}\label{sec:problem}

\begin{figure}
  \centering
  \includegraphics[width=0.85\textwidth]{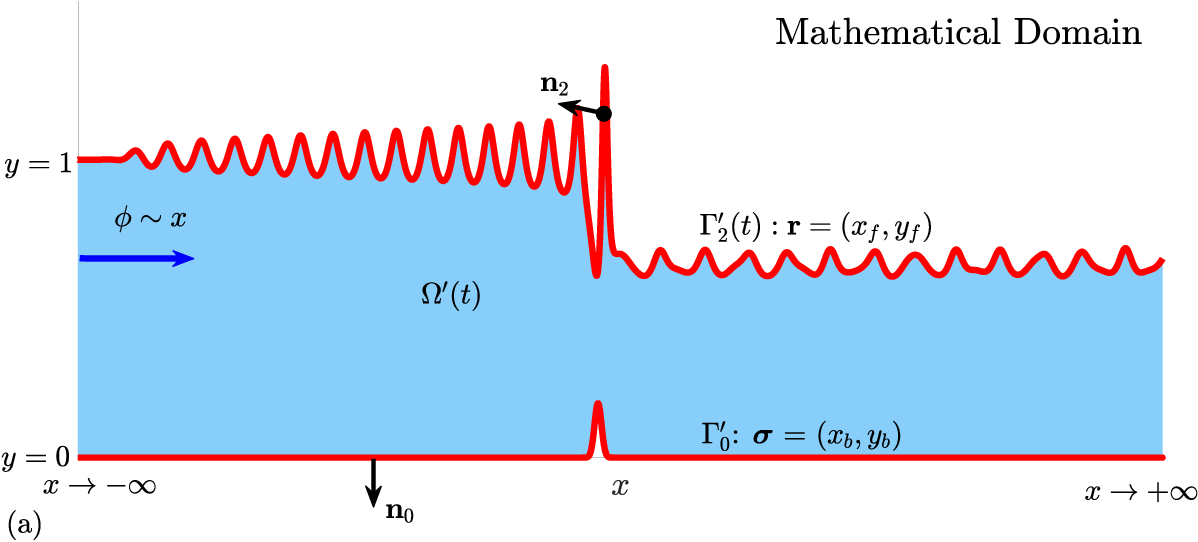}
  \includegraphics[width=\textwidth]{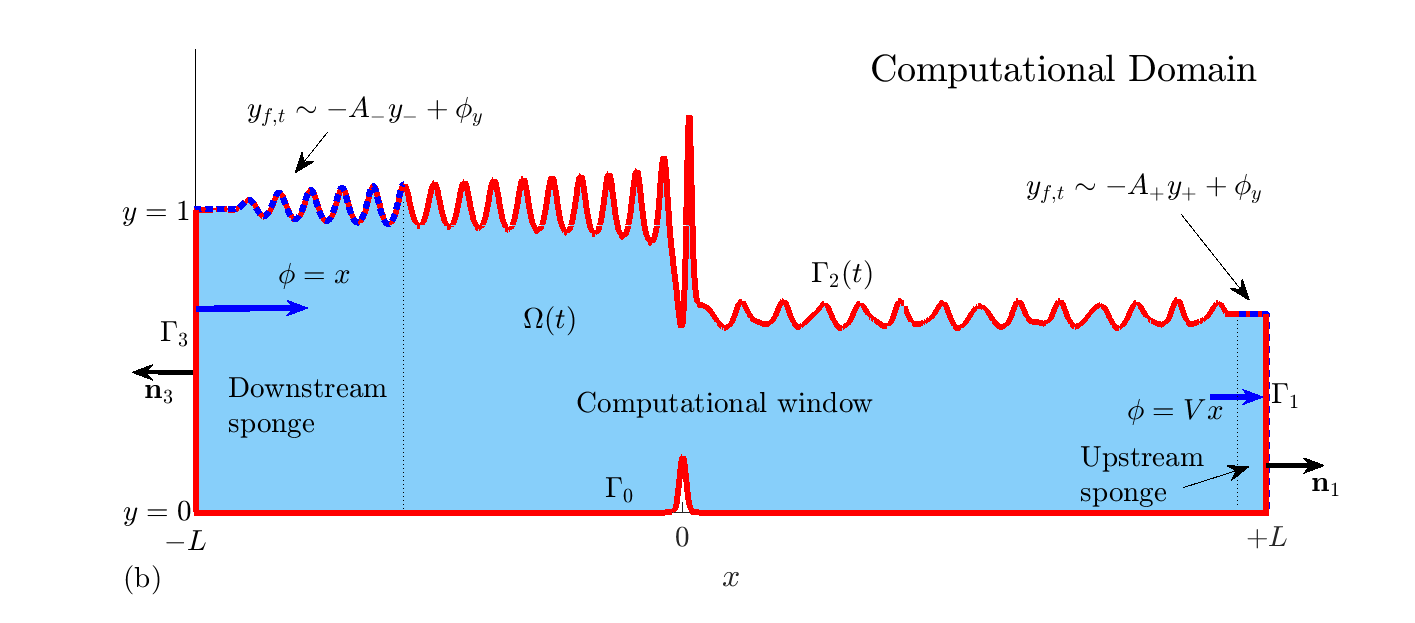}
  \caption{Sketch of the non-dimensional time-dependent problem domain for the hydraulic-fall problem. (a) The mathematical domain. The fluid domain is $\Omega'(t)$, the bottom boundary is $\Gamma'_0$ and the free-surface is $\Gamma'_2$. As $x\to-\infty$, we impose uniform flow; $\phi\sim x$. (b) The computational domain. The inflow boundary is denoted $\Gamma_3$ and the outflow boundary is $\Gamma_0$. The upstream and downstream sponges are indicated by dotted lines on the free-surface $\Gamma_2$. In the computational domain, far downstream we impose that $\phi=Vx$. In both panels, the normal vectors to each boundary $\Gamma_i$ are given as $\textbf{n}_i$.}
  \label{fig:problem_domain}
\end{figure}

We consider inviscid, irrotational, incompressible flow over a bottom topography under the influence 
of gravity and neglecting surface tension. The general flow scenario is depicted in 
figure~\ref{fig:problem_domain}. The fluid flows from left to right over a localised topographic structure seen as a positive bump in the figure. In the absence of the topography so that the bottom is flat, or sufficiently far upstream of the obstacle, the flow 
consists of a uniform stream of strength $U$ with uniform fluid depth $H$. Our concern is with 
the generally unsteady disruption that is provoked at the free surface by the topography.  

We solve for the fluid flow and the deformation of the free surface using a numerical finite-element method. In the following sub-sections we first describe the mathematical problem to be solved before discussing the truncated computational problem that is required for the numerical implementation. To aid the discussion, we use primes to denote regions and boundaries in the mathematical problem. For example, $\Omega'(t)$ represents the fluid domain in the mathematical problem, and $\Omega(t)$ represents the corresponding truncated computational domain.
%%%%%%%%%%%%%%%
%%%%%%%%%

\subsection{Mathematical formulation}\label{sec:math}

We non-dimensionalise the problem by scaling all lengths and velocities by the height $H$ and strength $U$ of the oncoming uniform stream, respectively.  The corresponding time scale is $H/U$. Figure~\ref{fig:problem_domain}(a) shows the domain for the mathematical problem expressed in terms of dimensionless variables and with reference to a Cartesian set of axes $(x,y)$. The free-surface is described by $\textbf{r}= (x_f,y_f)$ and the bottom topography by $\boldsymbol{\sigma} = (x_b,y_b)$. Written in terms of the velocity potential, $\phi$, the dimensionless governing equation and boundary conditions are
\begin{align}
  &\nabla^2\phi = 0,\quad \textbf{x}\in\Omega'(t)\quad &\mbox{(Conservation of mass)}\label{bulk_eq}\\
  &\nabla\phi\cdot\textbf{n}_0 = 0,\quad \textbf{x}\in\Gamma'_0,\quad&\mbox{(No penetration on bottom)}\label{bottom_eq}\\
  &\black{\pdiff{\textbf{r}}{t}\cdot\textbf{n}_2 = \nabla\phi\cdot \textbf{n}_2 ,\quad \textbf{x}\in\Gamma'_2(t),} \quad &\mbox{(Kinematic condition)}\label{kin_eq}\\
  &\phi_t + \frac{1}{2}|\nabla\phi|^2 + \frac{1}{Fr^2}\left(y_f-1\right) - \frac{1}{2} = 0\quad \textbf{x}\in\Gamma'_2(t).\: &\mbox{(Dynamic condition)}
  \label{dyn_eq}
\end{align}
The Froude number $Fr$ was defined in \eqref{froude}.
The dynamic boundary condition follows from applying Bernoulli's equation at the free surface and utilising the upstream uniform flow conditions, namely
\begin{align}
 & y_f\to 1,\: \phi \to x \qquad\mbox{as}\qquad x\to-\infty. %, \qquad &\mbox{(Uniform flow).}
  \label{flat}
\end{align}
Throughout we shall adopt the Gaussian topographic forcing function
\bea
y_b = a\mbox{e}^{-b^2x_b^2},
\label{sig}
\eea
for some $a,b\in\mathbb{R}$.
%Finally, as $x\to\infty$ we denote the flux Far downstream the flux is $q(t)$ so the total number of parameters that are required to be specified are $[Fr, a, b, q]$; how $q$ is determined depends on whether the calculation is steady or time-dependent as we now discuss.

\subsection{Truncated computational formulation}\label{sec:numform}

To prepare for the numerical discretisation, as shown in figure~\ref{fig:problem_domain}(b) we truncate the infinite domain so that ${x}\in[-L,L]$, with $L$ to be chosen to be sufficiently large. The problem on the truncated domain is given by \eqref{bulk_eq}--\eqref{dyn_eq} (with the primes on the flow domain and boundary symbols removed). Boundary conditions must be imposed at the artificial outflow and inflow boundaries, $\Gamma_1$ and $\Gamma_3$, respectively, that arise due to the truncation.

At the inflow boundary $\Gamma_3$, corresponding to ${x}=-L$, the velocity potential and free surface level are set to mirror the condition \eqref{flat} in the mathematical 
formulation. Accordingly, we impose
\bea
{y}_f = 1,\quad {\phi} = {x}, \qquad {\textbf{x}}\in\Gamma_3, \qquad (\mbox{Inflow condition).}
\label{in_num}
\eea
which imposes a Dirichlet condition on $\phi$.
The conditions to be imposed at the outflow boundary are more subtle. Anticipating the presence of waves propagating downstream, for our unsteady calculations we will make use of a sponge layer to force the free surface to become locally flat downstream allowing us to impose the outflow Neumann boundary condition
\bea
\nabla{\phi}\cdot\textbf{n}_1 = V \qquad\mbox{on}\qquad \textbf{x}\in\Gamma_1,\qquad (\mbox{Ouflow condition)}
\label{out_num}
\eea
where $\textbf{n}_1=\textbf{i}$ is the unit vector in the $x$ direction, and $V$ is set to ensure conservation of mass. In this way we expect to capture the genuine flow features across the main part of the computational domain (the `computational window' -- see figure~\ref{fig:problem_domain}b) except in a short region close to the outflow boundary where the flow is artificially forced into becoming a locally uniform stream. The choice of $V$ together with details of the sponge-layer will be discussed below.

\subsection{Steady calculations}\label{sec:steady_flux}

In presenting the steady solution space we 
include both types for completeness and to provide greater context.
Since hydraulic rise solutions are subcritical at the downstream end (contrast the hydraulic 
fall solutions sketched in figure~\ref{fig:summary}), we found that in practice it is convenient 
to replace the conditions on $\phi$ in \eqref{in_num} and \eqref{out_num} with a Neumann 
condition on $\phi$ at the inflow boundary $\Gamma_3$ and a Dirichlet condition on $\phi$ at 
the outflow boundary. In this way we can prevent the occurrence of cnoidal waves downstream,
 
 As was mentioned above, the downstream speed $V$ in \eqref{out_num} is chosen to ensure conservation of mass. If $\gamma_s = y_f(x=L)$ is the 
{\it a priori} unknown free surface level local to $\Gamma_1$, then we must choose $V=1/\gamma_s$. The steady form of the 
dynamic boundary condition \eqref{dyn_eq} then requires
\begin{align}
&\frac{1}{2}\gamma_s^{-2} + \frac{1}{Fr^2}\left(\gamma_s - 1\right) - \frac{1}{2} = 0.
\label{out_flux_eq}
\end{align}
This is satisfied for any $Fr$ if $\gamma_s=1$, but if $\gamma_s\neq 1$ it imposes a constraint on the Froude number, and thus the Froude number must come as part of the solution.

While our focus in this paper is on hydraulic fall solutions, we can also calculate hydraulic-rise solutions for which the fluid depth 
increases monotonically from its upstream level, where $Fr>1$, to its downstream level, where $Fr<1$. These solutions are not observed in experiments and so the main focus of our stability analysis is hydraulic falls, although we do present a brief description of the nonlinear stability of hydraulic rises for completeness.

%{\color{red} To be reviewed: We do not consider these types of solution here because the boundary conditions for the steady configuration are incompatible with the time-dependent boundary conditions, as shall be discussed in \S~\ref{sec:unsteady_flux}. }

%%%%%%%
\subsection{Time-dependent calculations}\label{sec:unsteady_flux}

In the time-dependent problem when waves reach the artificial boundaries in the truncated 
computational domain at $x=\pm L$ they will either be reflected back into the domain or the 
simulation will fail. A number of different approaches have been proposed for dealing with this 
issue (see \cite{romate1991} for a review). These include imposing appropriate radiation 
conditions \citep{buttle2018three,ctugulan2022three} or introducing perfectly-matched
layers (see, for example, \cite{bermudez2007optimal} for acoustic waves). We choose a 
simpler approach by introducing so-called sponge-layers adjacent to the inflow and outflow boundaries. The sponge-layer was described by \cite{boyd1} and \cite{alias}, and it has been implemented, for example, by \cite{grimshawsponge} and 
\cite{keelercriticalflow} in the case of the forced KdV equation.

In the sponge layers, disturbances near to the inflow and outflow boundaries are exponentially damped so that 
\begin{equation}
  y_f - y_{\pm}\propto \mbox{e}^{-A_{\pm}t},
  \label{sponge}
\end{equation}
where $y_-$ and $y_+$ are, respectively, target upstream and downstream free surface levels that will be discussed below, and $A_+$ and $A_-$ are specified decay rate constants. To achieve this we modify the kinematic condition, \eqref{kin_eq}, to become
\bea
\pdiff{\textbf{r}}{t}\cdot\textbf{n}_2 - \underbrace{S_-(x)(y_f - y_-)}_{\mbox{Upstream sponge}} - \underbrace{S_+(x)(y_f - y_+)}_{\mbox{Downstream sponge}} = \nabla\phi\cdot \textbf{n}_2.
\eea
The sponge functions $S_{\pm}(x)$ take the form
\bea
S_{+}(x) = A_+[1+\tanh(B_+(x - C_+))],\quad S_{-}(x) = A_-[1 + \tanh(-B_-(x - C_-))], 
\eea
so that the exponential damping only occurs over specified regions dictated by the parameters $B_{\pm}$ and $C_{\pm}$. Care must be exercised when choosing values for $A_{\pm},B_{\pm},C_{\pm}$ to ensure that free surface waves do not get reflected back into the main part of the computational domain. After extensive experimentation, we found that taking $A_{\pm}=5$ is appropriate, and setting $B_+ = 1$ ensures that small amplitude waves are absorbed downstream, and taking $B_-= 0.01$ avoids large amplitude waves being reflected back into the domain at the upstream end. We set the locations of the sponge layers by taking $C_{\pm} = \pm L \mp 10$, and typically we choose $L>100$ -- see figure~\ref{fig:problem_domain}(b) where the sponge region is marked approximately by dashed lines on $\Gamma_2$.

The upstream target level we set as $y_{-} = 1$, whereas for the downstream target level, we either set $y_{+} = \gamma_s$ or $y_{+} = 1$ 
depending on the initial condition. This choice will be discussed in \S~\ref{sec:nonlinear_stability} and \S~\ref{sec:unsteady_inverse}. Regardless 
of the choice of $\gamma_s$, we fix $V=1/y_f(L,t)$ so that the outflow flux is equal to unity. We stress that this technique of damping the waves 
upstream and downstream is independent of the underlying numerical discretisation of the system. It could 
in principle be applied to the analogous 3D problem as an alternative to the radiation conditions used in \cite{buttle2018three,ctugulan2022three}.

\subsection{Weak formulation}\label{sec:weak_form}

The governing equations, \eqref{bulk_eq}--\eqref{dyn_eq} are highly nonlinear and numerical methods are typically required to solve them. Boundary-integral methods are popular \citep[see, for example][]{binder+vanden,dias+vanden-broeck,binder2008,forbes1982,wade+binder,wade+binder2}, although finite-difference schemes have been used \citep{grimshaw1} and more recently a spectral method has been implemented \citep{forbes2021ideal}. We choose a different approach and develop a numerical framework based on a weak formulation of the governing equations over the computational domain (figure~\ref{fig:problem_domain}b). The advantages of this approach shall be discussed in \S~\ref{sec:num_dis} but first we describe the mathematical weak formulation.

We multiply \eqref{bulk_eq} by a test function, $\psi(\textbf{x})$, which is required to vanish on boundaries where Dirichlet conditions in $\phi$ are imposed. Integrating \eqref{bulk_eq} over the computational domain and integrating by parts yields
\bea
\iint_{\Omega(t)} \left(\nabla^2\phi\right)\,\psi\,\mbox{d}V \equiv \int_{\partial\Omega(t)}\left(\textbf{n}\cdot\nabla\phi\right)\psi\,\mbox{d}S-\iint_{\Omega(t)} \nabla\phi\cdot \nabla\psi\,\mbox{d}V = 0,
\label{femeq1}
\eea
where $\partial\Omega(t)$ represents the boundary of $\Omega(t)$, \textbf{n} is the outwards-pointing unit normal vector on each part of the boundary and $\mbox{d}V$, $\mbox{d}S$ are differential area and line elements, respectively. The domain boundary can be decomposed into $\partial\Omega(t) = \Gamma_0 + \Gamma_1 + \Gamma_2(t) + \Gamma_3$ and so, imposing the Neumann conditions \eqref{bottom_eq}, \eqref{kin_eq} and \eqref{out_flux_eq}, means that \eqref{femeq1} becomes
\bea
\mathcal{R}_{\mathrm{Bulk}}(\textbf{x},\phi(\textbf{x}))  \equiv   \int_{\Gamma_2(t)}\underbrace{\left(\pdiff{\textbf{r}}{t}\cdot\textbf{n}_2\right)}_{\mbox{Kinematic Cond.}}\psi\,\mbox{d}S +\int_{\Gamma_1}\underbrace{V}_{\mbox{Outflow}}\psi\,\mbox{d}S -\iint_{\Omega(t)} \nabla\phi\cdot \nabla\psi\,\mbox{d}V = 0,
\label{femeq3}
\eea
with the unit normal $\textbf{n}_2$ identified in figure~\ref{fig:problem_domain}. To satisfy the dynamic boundary condition we multiply \eqref{dyn_eq} by a test function and integrate over the free-surface to obtain 
\bea
\mathcal{R}_{\mathrm{Dyn}}(\textbf{x},\phi(\textbf{x}))\equiv\int_{\Gamma_2(t)}\left(\phi_t + \frac{1}{2}|\nabla\phi|^2 + \frac{1}{Fr^2}\left(y_f-1\right) - \frac{1}{2}\right)\psi\,\mbox{d}S = 0.
\label{dyn_weak_form}
\eea
This finalises the weak formulation of the problem. Equations \eqref{femeq3} and \eqref{dyn_weak_form} are to be solved to determine for the free-surface location, $y_f(x,t)$, and the velocity potential in the fluid, $\phi(\textbf{x},t)$.

\subsection{Numerical discretisation}\label{sec:num_dis}

%Typically for steady calculations a nonlinear iterative solver, for example Newton's method, is used for steady states and explicit time-stepping is used for time-dependent calculations. For 3D problems, a Green's function formulation has also been developed, \citep[see, for example][]{buttle2018three,ctugulan2022three}, but, despite only involving quantities on the free-surface, the numerical Jacobian is dense and hence pre-conditioning and iterative linear algebra routines are usually required. In this article we formulate a numerical finite-element method (FEM) framework, based on a variational formulation, to study hydraulic-fall solutions to the steady and time-dependent problem of the FNL. The FEM framework we develop here has a larger number of unknowns but the resulting Jacobian matrix is sparse and can be inverted efficiently with `out of the box' direct linear solvers such as LU factorisation \citep{li2005overview}.

To solve \eqref{femeq3} and \eqref{dyn_weak_form} numerically, we utilise the open-source C++ package \texttt{oomph-lib} \citep{heil2006oomph} that discretises the governing equations using a FEM Galerkin method and allows us to take advantage of state-of-the-art linear solvers, mesh-update techniques and provides flexibility to switch between steady, time-dependent and linear stability calculations. We utilise an isoparametric representation, wherein the same interpolating shape functions are used for $\psi$ and for the position variables. For this problem we used piecewise cubic shape functions and typically chose 400 $\times$ 10 elements in the $x\times y$ directions. We use a structured quadrilateral mesh with either a spine-node update strategy (for the stability analysis and time-dependent calculations) or a pseudo-solid elastic mesh update strategy (for the steady calculations), although in principle other mesh geometries, such as an unstructured triangular mesh, can be used. Newton iterations are used to calculate steady states and a backwards-difference Euler order 2 (BDF2) implicit method is used for unsteady time-stepping; after extensive experimentation we set the timestep $\Delta t = 1.0$, which is small enough to resolve the temporal features of the solutions to be discussed. An attractive feature of this formulation is that time-dependent calculations only require a very minor (and easy) augmentation of the Jacobian matrix used in the Newton iterations and so it is straightforward to switch between steady-state calculations and time-dependent calculations. Another notable feature is that despite calculating quantities in the bulk fluid as well as the free-surface, the corresponding Jacobian matrix is sparse, in contrast to the boundary-integral approaches, and the linear inversion is quick based on the open-source SuperLu package \citep{li2005overview} which performs LU factorisation. Finally, in the weak formulation, the numerical generalised eigenvalue problem that results from the linear stability analysis is highly rank deficient as time-derivatives do not explicitly appear in the bulk fluid. To compute the eigenvalues and eigenmodes accurately and efficiently, we implement the eigensolver from the Anasazi linear algebra library \citep{herouxtrilnos} that is based on Arnoldi-iteration and has been used successfully in other rank-deficient generalised eigenproblems \citep{thompson2013stability,keeler2019invariant}. %\past{We direct the reader to XXX where they can download the user files that solve the steady problem, linear stability analysis and time-dependent calculations.}\textbf{\past{JSK: I'll put the driver codes on figshare and then put a link on the data availibility section at the end}}.

%%%%%%%%%%%
\section{Steady solutions, stability analysis, and time-dependent simulations}\label{sec:stability}

In this section we first describe the steady hydraulic-fall solution structure. Next we establish the linear stability properties of the hydraulic-fall solutions by solving a generalised eigenvalue problem and probe the nonlinear stability properties by performing time-dependent simulations.

%%%%%%%
\subsection{Steady bifurcation structure}\label{sec:fnl}

%%%%%%%%%%
\begin{figure}
  \centering
  \includegraphics[width=\textwidth]{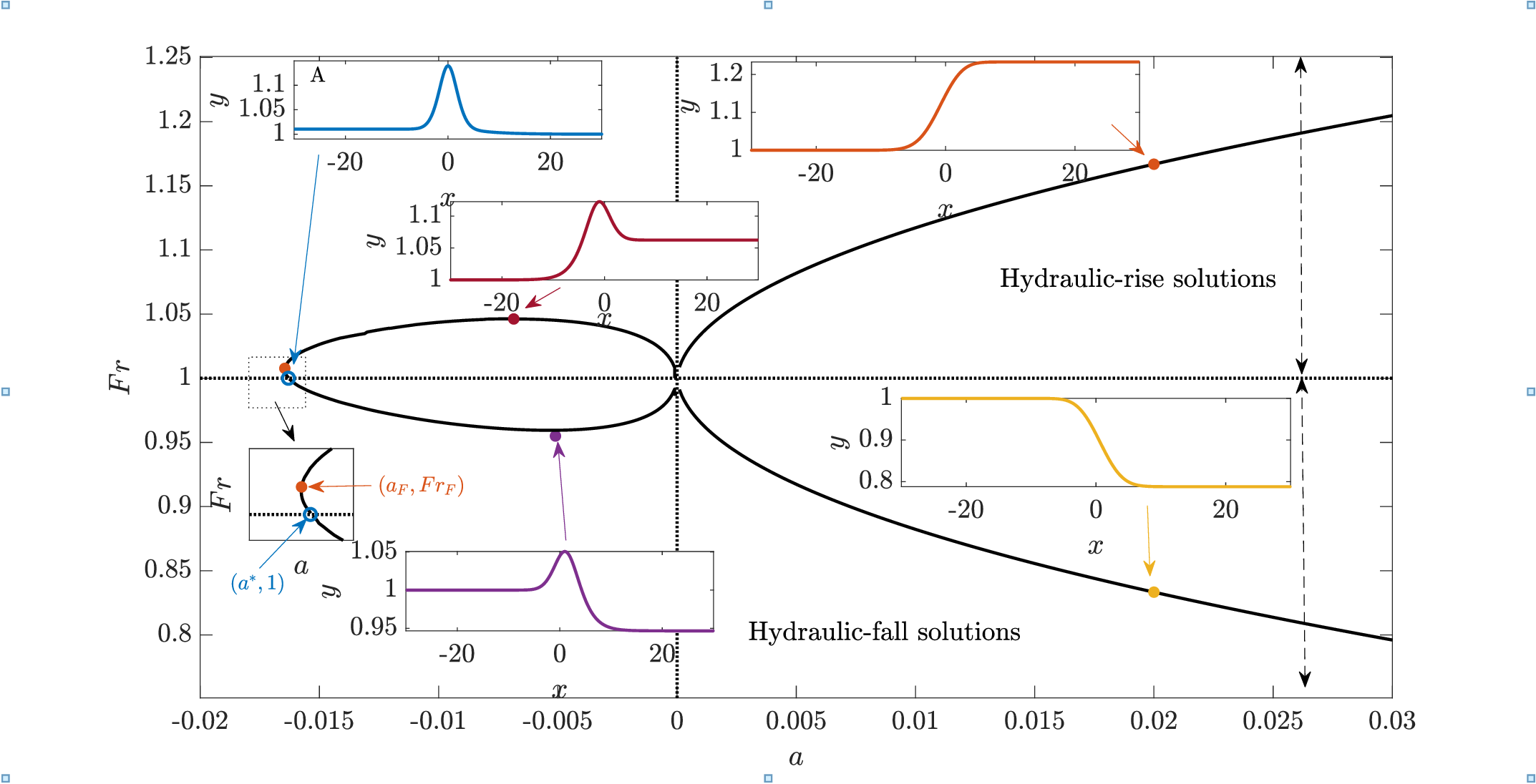}
  \caption{Steady solution space for $b = 0.3$. The amplitude of the forcing, $a$, is the horizontal measure and the vertical measure is the Froude number, $Fr$. The inset panels show FNL profiles on the branch indicated by arrows, Here $a^*\approx  -0.016$, $(a_F, \, Fr_F) \approx (-0.0165,1.006)$.}
  \label{fig:fnl_hfall}
\end{figure}
%%%%%%%%%%

We briefly describe the steady bifurcation structure for the FNL system. In 
figure~\ref{fig:fnl_hfall} we show the hydraulic-fall solution space in the $(a,Fr)$ plane. The 
hydraulic-rise steady states are included for completeness.

We focus on the difference between the free-surface profiles for $a<0$ and $a>0$. For 
$a>0$ the hydraulic-fall/rise profiles are monotonic decreasing/increasing functions of 
$x$, respectively, and there is a unique hydraulic fall solution for each $a$. For $a<0$ the profiles are non-monotonic and in each case the wave height reaches a maximum close to $x=0$ before monotonically decreasing (vice-versa for the hydraulic-rise profiles). 

Unlike the corresponding curve for the delta-function-forced KdV equation \cite[see, 
for example][]{binderreview}, the solution curve in figure~\ref{fig:fnl_hfall} is neither symmetric about $a=0$ nor about $Fr=1$ (although the latter appears true, close inspection reveals this not to be the case -- see the zoomed inset diagram in the figure). 
Again in contrast to what is observed for the forced KdV equation, the solution curve has a turning point which is located at $(a_F,Fr_F)$, with $Fr_F>1$, so that there is a parameter window, $a_F< a < a^*$, in which there is a multiplicity of  
hydraulic-rise solutions.
% In the hydraulic-fall section of the curve there is no such turning point. 

There are no hydraulic fall solutions for $a \leq a^*$. We encounter numerical difficulties in 
reaching the point $a=a^*$. As the branch approaches this point the wave-profile, as shown 
in the top left inset of figure~\ref{fig:fnl_hfall} (labelled A) appears to approach a solitary-wave type 
solution in which the downstream height $\gamma_s\to 1$. We believe that the profile at 
$a=a^*$ corresponds to that at the termination point of the $Fr=1$ branch of solution (not computed here), as 
described by \cite{keelernonlinear} for the fKdV model.
%which is in agreement with the results of \cite{ee+grimshaw} using an fKdV model. As this point is approached from above, $\gamma_s$ tends to 1 and we would expect that the limiting profile to be a solitary-wave solution. 

\subsection{Linear stability analysis}\label{sec:linear_stability}

To analyse the stability of the steady solutions discussed in the previous section, we first introduce the velocity potential on the free-surface, $\varphi(x,t) \equiv \phi(x,y_f,t)$, and the surface elevation $\eta \equiv y_f - 1$. We then write
\bea
\varphi = \varphi_s(x) + \hat{\varphi}(x,t),\qquad  \eta = \eta_s(x) + \hat{\eta}(x,t),
\label{base_pert}
\eea
where $\varphi_s(x)$ and $\eta_s(x)$ are the velocity potential and elevation for a steady hydraulic fall solution, and the hatted variables are time-dependent perturbations which are assumed to vanish at infinity. We emphasise that the perturbations are arbitrary and not at this point assumed to be small.

It is helpful for the subsequent linear stability analysis to write the system of equations in Hamiltonian form. However, since we perturb about steady hydraulic-fall solutions for which the 
surface potential and elevation functions, $\varphi_s$ and $\eta_s$, do not vanish as $x\to \infty$, some modification to the well-known Craig-Sulem-Zakharov (CSZ) Hamiltonian formulation of the 
water wave problem \citep{zakharov1968stability,craig1993numerical} is necessary, which we now describe. We note that the base state in \eqref{base_pert} does not necessarily have to represent a hydraulic-fall solution; the following analysis is also valid if the underlying steady state is a hydraulic-rise or solitary-wave solution, for example.

%\subsubsection{Construction of the Hamiltonian} \label{app:hamiltonian}

In constructing the Hamiltonian it is important to realise that \eqref{base_pert} reflects a nonlinear perturbation to free-surface quantities, and that some care must be exercised when defining corresponding bulk velocity potentials. In particular we define $\xi(x,y,t)$ and $\hat{\phi}(x,y,t)$ to be the solutions to the following Dirichlet-Neumann problems, denoted A and B, which are both defined on the time-dependent domain $\Omega'(t)$:
\bea
\mbox{Problem A}:\left\{
\begin{split}
  \nabla^2 \xi = 0\qquad \mbox{in} \qquad \Omega'(t),\nonumber \\
  \xi = \varphi_s\qquad \mbox{on} \qquad \Gamma'_2(t),\nonumber \\
  \textbf{n}_0\cdot\nabla \xi = 0\qquad \mbox{on} \qquad \Gamma'_0,\nonumber\\
  \xi \sim x \quad \mbox{as} \quad x \to -\infty,\nonumber\\
  \xi \sim \gamma_s^{-1}x  \quad \mbox{as} \quad x \to \infty,\nonumber\\
\end{split}
\:\right. \quad
\mbox{Problem B}:\left\{
\begin{split}
  \nabla^2 \hat{\phi} = 0\qquad \mbox{in} \qquad \Omega'(t),\nonumber \\
  \hat{\phi} = \hat{\varphi}\qquad \mbox{on} \qquad \Gamma'_2(t),\nonumber \\
  \textbf{n}_0\cdot\nabla \hat{\phi} = 0\qquad \mbox{on} \qquad \Gamma'_0,\nonumber \\
  \hat{\phi} \to 0 \quad \mbox{as} \quad x \to -\infty,\nonumber \\
  \hat{\phi} \to 0 \quad \mbox{as} \quad x \to \infty.
\end{split}
\:\right.
\eea
We assume $\hat{\eta}$ and higher-order corrections to $\xi$ decay sufficiently fast as $x\to\pm\infty$ for the integrals that we will state below (e.g. \eqref{hamiltonian_base}) to be convergent. A subtle point to note is that $\xi$ does not correspond to the velocity potential of the steady state as it is defined on the time-dependent domain, which in general does not coincide with that for the steady state. However, the trace of $\xi$ on the free-surface coincides with that for the steady problem (we note that this construction relies on the assumption that the free-surface is a graph).

The combined velocity potential $\phi = \xi + \hat{\phi}$ 
satisfies Laplace's equation, \eqref{bulk_eq}, in the time-dependent domain. Taking inspiration from the CSZ construction, Problems A and B have been formulated in such a way that we may write down a Hamiltonian for our problem in the form
\begin{equation}
  \begin{split}
    \hat{\mathcal{H}}(\hat{\varphi},\hat{\eta}) =  \hat{\mathcal{H}}_{0}(\xi,\eta_s,\hat{\eta}) + \underbrace{\frac{1}{2}\iint_{\Omega'(t)} |\nabla\hat{\phi}|^2\,\mbox{d}V + \frac{1}{2Fr^2}\int_{-\infty}^{\infty}\hat{\eta}^2\,\mbox{d}x}_{\mathrm{CSZ}}\\+\underbrace{\iint_{\Omega'(t)}\nabla\xi\cdot\nabla\hat{\phi}\,\mbox{d}V +\frac{1}{Fr^2}\int_{-\infty}^{\infty}\eta_{s}\hat{\eta}\,\mbox{d}x}_{\mathrm{Perturbation}}.
  \end{split}
  \label{hamiltonian}
\end{equation}
We have split $\hat{\mathcal{H}}$ into a term that replicates the original CSV Hamiltonian, a perturbation term that arises from the nonlinear perturbation in \eqref{base_pert}, and a term that can be considered as the `base' energy, $\hat{\mathcal{H}}_0$, defined as
\begin{equation}
  \begin{split}
      \hat{\mathcal{H}}_{0}(\xi,\eta_s,\hat{\eta}) = \int_{-\infty}^{\infty}\left[\frac{1}{2}\int_{0}^{1+\eta_s+\hat{\eta}}|\nabla\xi|^2 - 1\,\mbox{d}y + \frac{1}{2Fr^2}\eta_s^2 + Q\right]\,\mbox{d}x,\quad Q = \frac{1}{2Fr^2}\eta_s\left(\eta_s + 2\right).
  \end{split}
  \label{hamiltonian_base}
\end{equation}
It is important to note that the integral in \eqref{hamiltonian_base} is convergent due to the addition of the term, $Q$ (Since $Q$ does not depend on $\hat \phi$ or $\hat \eta$ it does not contribute on taking variations of $\hat{\mathcal{H}}$ with respect to these variables). As $x\to -\infty$ the terms inside the square brackets in \eqref{hamiltonian_base} vanish 
due to the inflow conditions \eqref{in_num}. As $x\to\infty$ they vanish due to the downstream condition \eqref{out_flux_eq}, posed on the mathematical domain as $x\to\infty$, 
assuming that we restrict the flow to approach a uniform stream with free-surface height either 1 or $\gamma_s$. 
This restriction as $x\to\infty$ is satisfied when discussing computations for nonlinear stability in 
\S~\ref{sec:nonlinear_stability}.

Finally, by using variational arguments as described in \cite{zakharov1968stability} and \cite{craig1993numerical} it can be shown that the system in \eqref{bulk_eq}--\eqref{dyn_eq}, together with \eqref{base_pert} can be written as
\bea
\frac{\partial\boldsymbol{\zeta}}{\partial t} = \textbf{K} \frac{\delta\hat{\mathcal{H}}}{\delta{\boldsymbol{\zeta}}},\qquad \textbf{K} = \begin{pmatrix} 0 & I\\ -I & 0 \end{pmatrix}, \qquad \frac{\delta}{\delta\boldsymbol{\zeta}} = \left(\frac{\delta}{\delta\hat{\eta}},\frac{\delta}{\delta\hat{\varphi}}\right)^T,
\label{dynamical_system}
\eea
where $\boldsymbol{\zeta} = (\hat{\eta},\hat{\varphi})^T$, where $\delta/\delta\boldsymbol{\zeta}$ is the variational derivative.  %\past{which we describe in appendix~\ref{app:hamiltonian}.}

Although the stability theory of Hamiltonian systems is well known \citep[see, for example][]{holm1985nonlinear}, it is helpful to repeat the salient details. First, we linearise \eqref{dynamical_system} about $\hat{\varphi}\equiv 0 \equiv \hat{\eta}$, that is we set $\boldsymbol{\zeta} = \varepsilon \bar{\boldsymbol{\zeta}}$, with $0<\varepsilon \ll 1$ and $\bar{\boldsymbol{\zeta}} = (\bar{\eta},\bar{\varphi})^T$. At $O(\varepsilon)$ we find
 \bea
  \pdiff{\bar{\boldsymbol{\zeta}}}{t} = \textbf{K}\textbf{L}\:\bar{\boldsymbol{\zeta}}, \qquad \textbf{L} = \frac{1}{2}\frac{\delta}{\delta\bar{\boldsymbol{\zeta}}}(\delta^2\hat{\mathcal{H}}),
  \label{pert_equation}
  \eea
  where $\delta^2\hat{\mathcal{H}}$ is the symmetric Hessian matrix (see, for example \cite{holm1985nonlinear}). Proceeding further, we write $\bar{\boldsymbol{\zeta}} = \textbf{g}_s\mbox{e}^{st}$, where $\textbf{g}_s = (g_{\hat{\eta},s}(x),g_{\hat{\varphi},s}(x))^T$ so that \eqref{pert_equation} becomes the eigenvalue problem
  \bea
  s\textbf{g}_s = \textbf{K}\textbf{L}\:\textbf{g}_s
  \label{evalue_equation}
  \eea
  for eigenvalues $s$ and eigenmodes $\textbf{g}_s$. 

  We distinguish between two classes of solutions to \eqref{evalue_equation} which correspond to i) a continuous essential spectrum, $s_{\mathrm{ess}}$, and ii) a discrete point spectrum, $s_{\mathrm{p}}$. The eigenmodes of $s_{\mathrm{ess}}$ are bounded as $|x|\to\infty$ whilst the eigenmodes of $s_{\mathrm{p}}$ decay to zero as $|x|\to\infty$. With $\textbf{K}$ skew-symmetric and assuming $\textbf{L}$ self-adjoint, it is easy to show that if $s=\rho\in\mathbb{C}$ satisfies \eqref{evalue_equation}, then so does $s=-\rho$ and $s=\pm\rho^*$ (stars here indicate complex conjugates) and therefore there is a four-fold symmetry in the complex plane. This is a standard distinguishing feature of Hamiltonian systems \citep[see, for example][]{holm1985nonlinear}.

  We calculate $s_{\mathrm{ess}}$ and $s_{\mathrm{p}}$ numerically using our finite-element framework (this implementation is identical to the procedure described in \cite{thompson2013stability}). The calculation is delicate and requires care to ensure convergence. In addition, we remark that to calculate $s_{\mathrm{ess}}$ and $s_{\mathrm{p}}$ we have to solve three separate numerical problems (one for $s_{\mathrm{ess}}$ and two for $s_{\mathrm{p}}$); the precise details will be discussed in parallel with the results below.
  
\subsubsection{The essential spectrum, $s_{\mathrm{ess}}$}
  
The essential spectrum, $s_{\mathrm{ess}}$, can be calculated by examining the form of 
\eqref{evalue_equation} in the limit as $|x|\to\infty$ \citep[see, for example][]
{sandstede2000absolute}. An interesting aspect of this problem is that the operator $\textbf{L}$ takes different 
forms as $x\to-\infty$ and as $x\to\infty$, and consequently different dispersion relations are 
obtained in these limits. This has implications for the spatial wave number of the 
eigenmodes.
%%%%%%%%
\begin{figure}
  \centering
  \includegraphics[width=\textwidth]{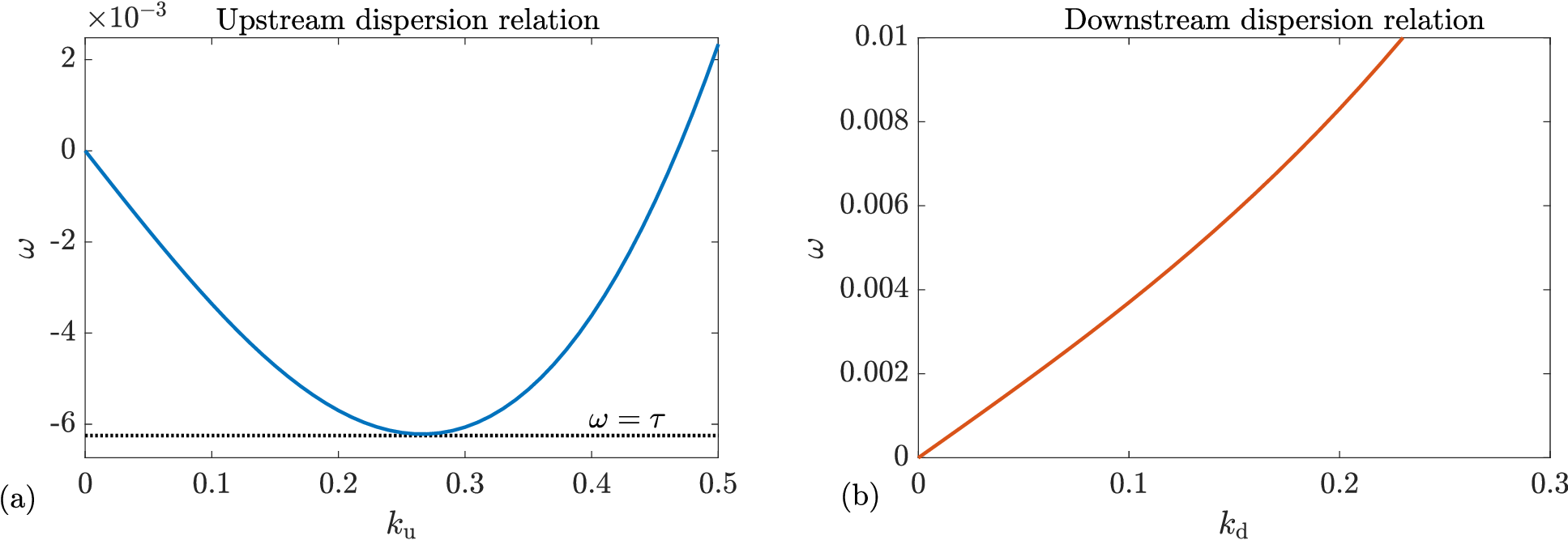}
  \caption{The dispersion curves for $Fr=0.9659$ and $\gamma_s = 0.9550$ (corresponding to the hydraulic-fall solution when $a=-0.01$, $b=0.3$). Panel (a): the upstream wave frequency $\omega$ as a function of the upstream wave number $k_{\mathrm{u}}$. Panel (b): the downstream wave frequency $\omega$ as a function of the upstream wave number $k_{\mathrm{d}}$. The minimum, $\omega = \tau = -0.006217$, of the upstream dispersion curve is shown with a dotted line in panel (a).}
  \label{fig:dispersion_curve}
\end{figure}
%%%%%%%

Sufficiently far downstream the flow corresponds to a uniform stream of height  $\gamma_s$ with speed $V = \gamma_s^{-1}$. The dispersion relation relating the frequency, $\omega$, to the wave number, $k_d$, of small amplitude waves is
 \bea
  \omega = Vk_{\mathrm{d}} \pm \sqrt{\frac{k_{\mathrm{d}}\tanh{k_{\mathrm{d}}\gamma_s}}{Fr^2}}, \qquad \mbox{(Downstream dispersion relation)}
  \label{downstream_disp}
  \eea
corresponding to waves which travel faster/slower, respectively, than the downstream speed, $V$. Upstream where the fluid depth and speed are both unity, with an obvious shift in notation the dispersion relation is
  \bea
  \omega = k_{\mathrm{u}} \pm \sqrt{\frac{k_{\mathrm{u}}\tanh{k_{\mathrm{u}}}}{Fr^2}}. \qquad \mbox{(Upstream dispersion relation)}
  \label{upstream_disp}
\eea
In both cases the essential spectrum is such that
\bea
s_{\mathrm{ess}} = \mathrm{i}\omega,\qquad \textbf{g}_{\mathrm{s}}\propto (\mathrm{e}^{\mathrm{i}k x},\mathrm{i}\mathrm{e}^{\mathrm{i}k x})^T,
\label{sess}
\eea
with $k=k_{u}$ or $k_d$, and $\omega \in \mathbb{R}$.
The profile of $g_{\hat{\eta},s}$ will be such that it connects a spatially oscillatory wave comprising wave numbers $k_{\mathrm{d}}$ that satisfy \eqref{downstream_disp} as $x\to \infty$, to a different spatially oscillatory wave consisting of wave numbers $k_{\mathrm{u}}$ that satisfy \eqref{upstream_disp} as $x\to -\infty$.

The dispersion relations in \eqref{downstream_disp} and \eqref{upstream_disp} are shown in 
figure~\ref{fig:dispersion_curve} for $a=-0.01,Fr=0.9659$. We plot the dispersion curves for positive $k$ and for the minus sign in \eqref{downstream_disp} and \eqref{upstream_disp}, but note that the dispersion curve for the plus sign can be obtained by simply replacing $k$ and $\omega$ with $-k$ and $-\omega$. We emphasise an important 
distinction between the upstream and downstream dispersion curves. As can be seen in panel 
(a) the upstream dispersion curve has a stationary point at a frequency which we denote $
\tau$, whereas the downstream dispersion curve monotonically increases with $k_{\mathrm{d}}
$, as can be seen in panel (b). This observation will be important when we describe the 
construction of $s_{\mathrm{ess}}$ below. 

To calculate $s_{\mathrm{ess}}$ numerically, we set $\phi = -L$ at the upstream boundary 
$\Gamma_3$ and allow the upstream height to come as part of the solution (in so doing we 
are able to capture waves upstream), and we `pin' the downstream 
height at the outflow boundary $\Gamma_1$.  Since we do not constrain $\phi$ at the outflow 
boundary this allows us to capture waves downstream. We filter out spurious essential modes 
by removing modes which have $|\mbox{Re}(s_{\mathrm{ess}})| > 10^{-7}$. We 
note that we do not apply the sponge layer in this calculation, i.e. we set $A_{\pm} = 0$.

\begin{figure}
  \centering
  \includegraphics[width=\textwidth]{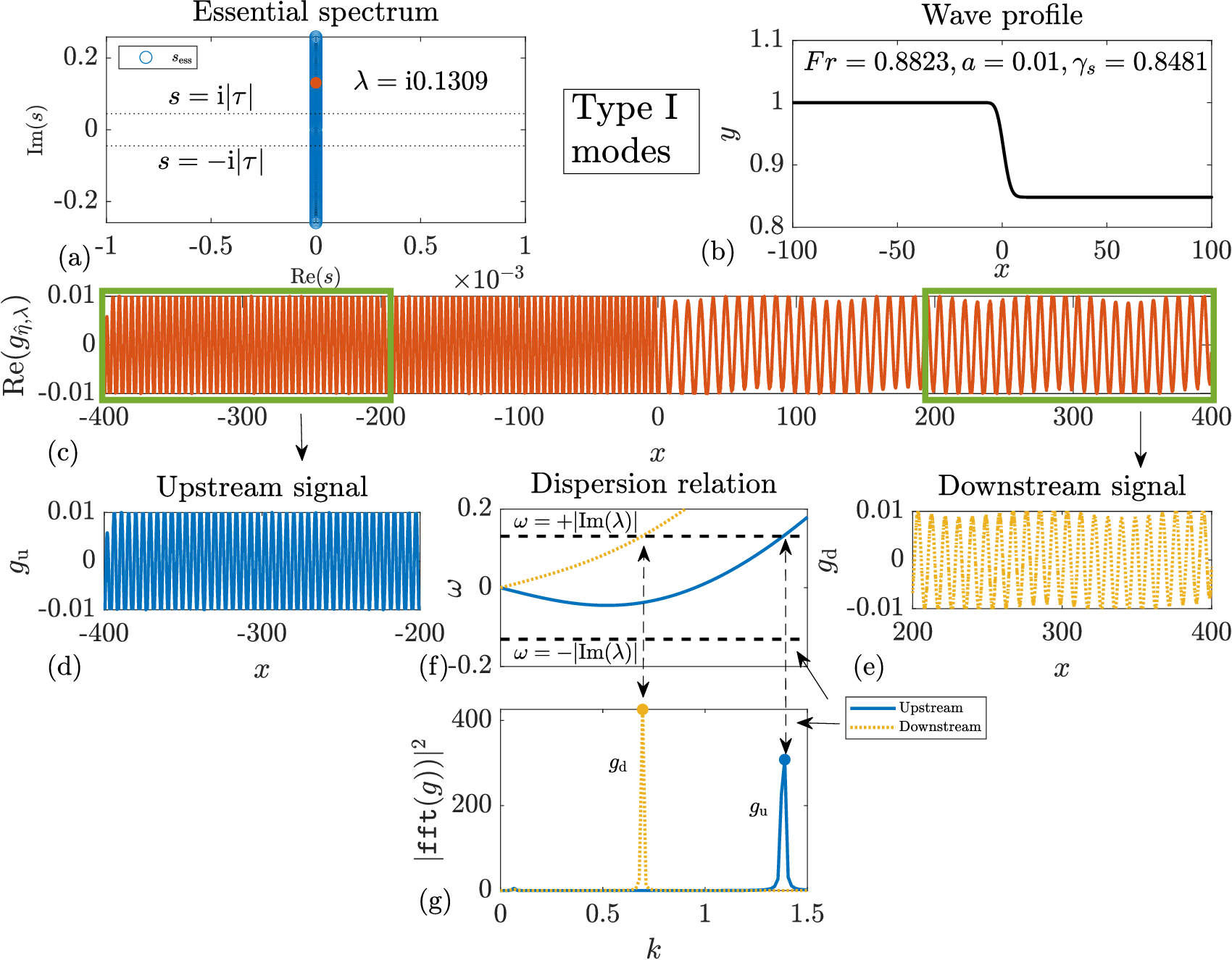}
  \caption{Type \soltype{1} modes of the numerically calculated $s_{\mathrm{ess}}$ for the hydraulic-fall solution, $Fr=0.8823$, $a=0.01$, $b=0.3$. Panel (a): The numerically computed $s_{\mathrm{ess}}$ is shown with blue markers on the imaginary $s$ axis, and a particular element of $s_{\mathrm{ess}}=\lambda=0.1309\mathrm{i}$ is highlighted with a solid red marker. The horizontal dotted lines indicate the levels $s=\pm \mathrm{i}|\tau| = \pm 0.04489\mathrm{i}$; in this calculation $|\mathrm{Im}(\lambda)| > |\tau|$. Panel (b): The underlying steady state. Panel (c): The real part of the eigenmode associated with $\lambda$. Panels (d), (e): The upstream/downstream portion of the real part of the eigenmode, denoted $g_{\mathrm{u}}/g_{\mathrm{d}}$, respectively. Panel (f): The downstream and upstream dispersion relation given in \eqref{downstream_disp}, \eqref{upstream_disp} respectively with the minus sign taken in both cases. The dashed horizontal lines indicate $\omega = \pm|\mathrm{Im}(\lambda)|$. Panel (g): The power spectrum, i.e. $|\texttt{fft}(g)|^2$ the square of the Fourier transform. The horizontal axes of panels (f) and (g) are identical so a direct comparison between the peaks of the power spectrum and the intersection of the dispersion curves with $\omega = \mathrm{Im}(\lambda)$ can be made. We also note that in panels (d)-(g) all calculations corresponding to the upstream section are shown as solid blue lines while the downstream section are dotted yellow lines.}
  \label{fig:alp_0_01_bi_modal}
\end{figure}

The essential spectrum comprises two types of modes, denoted types \soltype{1} and 
\soltype{2}. Type \soltype{1} modes occur when $|\mathrm{Im}(s_{\mathrm{ess}})| > |\tau|$. A 
set of calculations associated with a typical member of this class is shown in 
figure~\ref{fig:alp_0_01_bi_modal}. In panel (a) $s_{\mathrm{ess}}$ is shown on the imaginary 
axis in the complex $s$-plane for the underlying steady-state shown in panel (b) for 
$a=0.01,Fr=0.8823$. In panel (a) we highlight a particular member of $s_{\mathrm{ess}}$, 
denoted $\lambda$, which satisfies $|\mathrm{Im}(\lambda)| > |\tau|$ (values quoted in the 
caption). The corresponding mode (real part) is shown in panel (c) and comprises two distinct 
wave patterns upstream and downstream, denoted $g_{\mathrm{u}}$ and $g_{\mathrm{d}}$ 
and highlighted in panels (d) and (e) respectively. The dominant wave numbers of the type 
\soltype{1} modes can be determined, numerically, by calculating the power spectra of 
$g_{\mathrm{u}}$ and $g_{\mathrm{d}}$ as a function of the wave number $k$, as shown in 
panel (g). We calculate the power spectra using the fast Fourier transform (\texttt{fft}) of $g_{\mathrm{u}}$ 
and $g_{\mathrm{d}}$. For type \soltype{1} modes there is a single dominant wave number for 
each of the upstream and downstream signals as shown by the peak in each of the power 
spectra in panel (g). The wave numbers associated with these peaks correspond precisely to 
the intersections of the upstream and downstream dispersion curves with the horizontal line $
\omega = +|\mathrm{Im}(\lambda)|$, as shown in panel (f), which has the same horizontal axis 
and scale as panel (g) to aid this comparison. This illustrates the excellent agreement between 
the theory and numerics and gives us confidence in the fidelity of the numerical eigensolver.

\begin{figure}
\centering
\includegraphics[width=\textwidth]{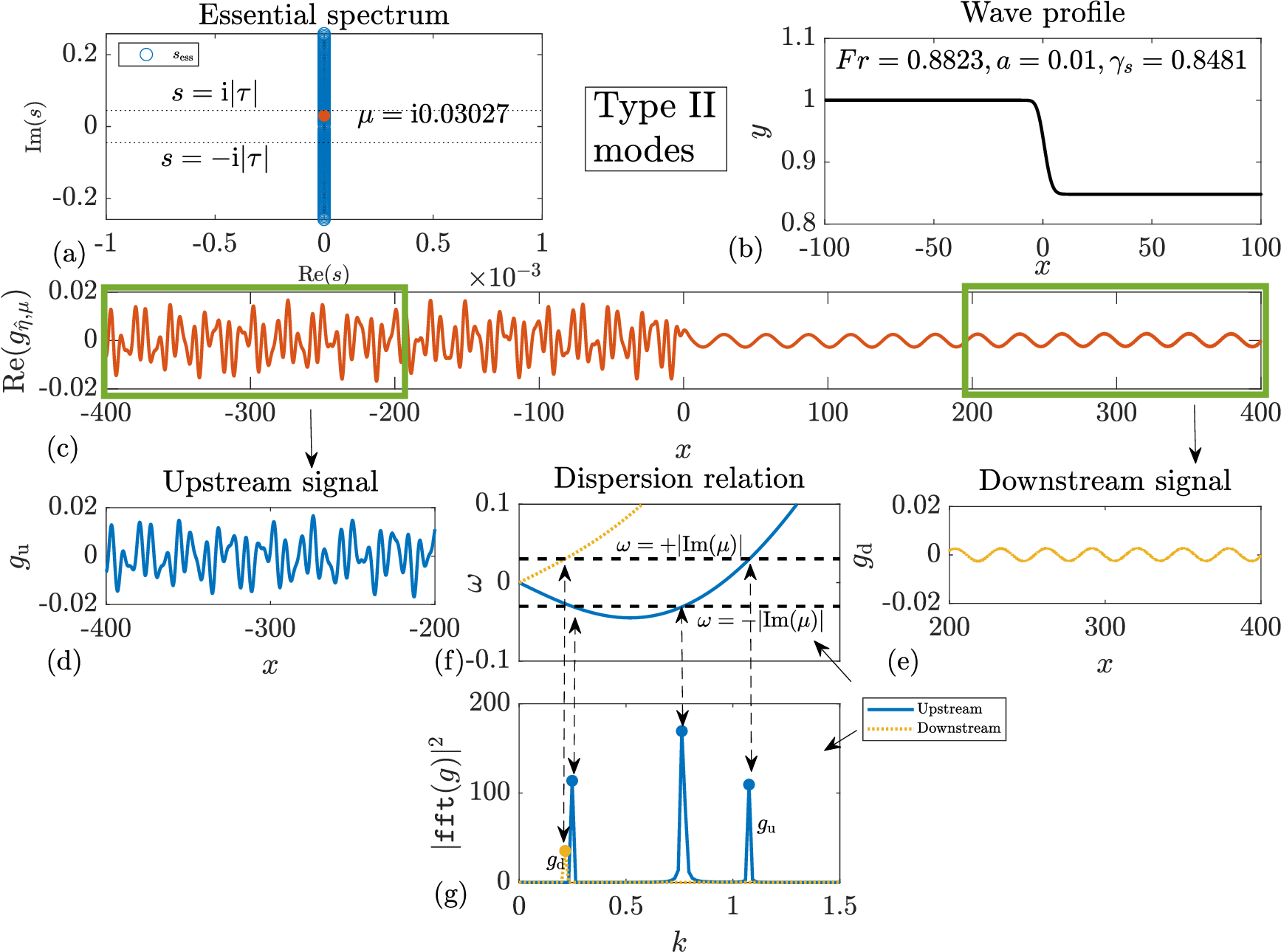}
\caption{Type \soltype{2} modes of the numerically calculated $s_{\mathrm{ess}}$ for 
the hydraulic-fall solution with $Fr=0.8823$, $a=0.01$, $b=0.3$. The figure follows the same 
description as figure~\ref{fig:alp_0_01_bi_modal} with the exception that in this calculation we 
highlight a different member of $s_{\mathrm{ess}}=\mu=0.03027\mathrm{i}$ such that 
$|\mathrm{Im}(\mu)| < |\tau|$. Here $\tau$ has the same value as in figure~\ref{fig:alp_0_01_bi_modal}.}
  \label{fig:alp_0_01_quad_modal}
\end{figure}

Type \soltype{2} modes occur when $|\mathrm{Im}(s_{\mathrm{ess}})| < |\tau|$. A set of 
calculations associated with a typical member of this class, denoted by $s_{\mathrm{ess}}
=\mu$, is shown in figure~\ref{fig:alp_0_01_quad_modal}. This figure follows an identical 
structure to figure~\ref{fig:alp_0_01_bi_modal}. In this class of modes, the upstream section of 
the mode, shown in panel (d), is clearly multi-harmonic. This is a direct 
consequence of the fact that in the upstream dispersion curve, for a given $|\omega| < |\tau|$, 
multiple values of $k_{\mathrm{u}}$ satisfy \eqref{upstream_disp}. For type \soltype{2} modes 
there are three dominant wave numbers upstream and one dominant wave number 
downstream. As for the type \soltype{1} modes, the wave numbers associated with the 
peaks in the power spectra correspond precisely to the intersections of the upstream and 
downstream dispersion curves, but this time with the horizontal lines $\omega = \pm|
\mathrm{Im}(\mu)|$, as shown in panels (f) and (g). For type \soltype{2} modes the wave 
numbers associated with $\omega = -|\mathrm{Im}(\mu)|$ will travel (as $t$ progresses) slower than the uniform upstream speed while wave numbers associated with $\omega = +|\mathrm{Im}(\mu)|$ will travel faster. 

%%%%%%%%%
%%%%%%%%%
%%%%%%%%%%%

\subsubsection{The point spectrum, $s_{\mathrm{p}}$}\label{sec:pointspec}
  
% For travelling-wave solutions to a one-dimensional evolution PDE (partial differential equation) $s_{\mathrm{p}}$ can be studied analytically and numerically using the Evans function \citep[see, for example][]{barker2017}. Adapting this methodology to two-dimensional problems is a considerable task so we calculate these numerically, directly using our finite-element framework as described below.

Throughout the following discussion we will use $\nu$ to represent the eigenvalue in the point spectrum, $s_{\mathrm{p}}$, that lies in the first quadrant of the complex plane. As was mentioned earlier, the point spectrum has a four-fold symmetry in the complex plane. In contrast to the forced solitary-wave fKdV calculations of \cite{camassa} and \cite{keelercriticalflow}, where the point spectrum eigenmodes are even-symmetric so that $g_{\hat{\eta},\nu}(-x) = g_{\hat{\eta},-\nu}(x)$, hydraulic-fall solutions are not even-symmetric and so we expect that $g_{\hat{\eta},\nu}(-x)\neq g_{\hat{\eta},-\nu}(x)$.

The numerical computation of the eigenmodes requires a great deal of care to filter out grid-dependent spurious modes.
A very wide computational domain is needed because evanescent waves on the upstream or downstream side of the eigenmodes decay 
extremely slowly (this is quantified below). We identify two different types of 
mode in the point spectrum which we term type \soltype{3} corresponding to eigenvalues in the left-half plane (here 
$s_{\mathrm{p}} = -\nu,\, -\nu^*$), and type \soltype{4} corresponding to eigenvalues in the right-half plane (here $s_{\mathrm{p}} = \nu,\, 
\nu^*$). To compute type \soltype{3} modes, we impose $y_f = \gamma_s$ and $\phi(L,y) = L$ at the outflow boundary $\Gamma_1$ so 
that the profile is flat downstream, but there are evanescent waves upstream. 
For type \soltype{4} modes, we set $y_f=1$ and $\phi(-L,y) = -L$, at the inflow boundary $\Gamma_3$ to capture eigenmodes that 
are flat upstream and have evanescent waves downstream. 

\begin{figure}
  \centering
  \includegraphics[trim = 0 70 0 0,clip,width=\textwidth]{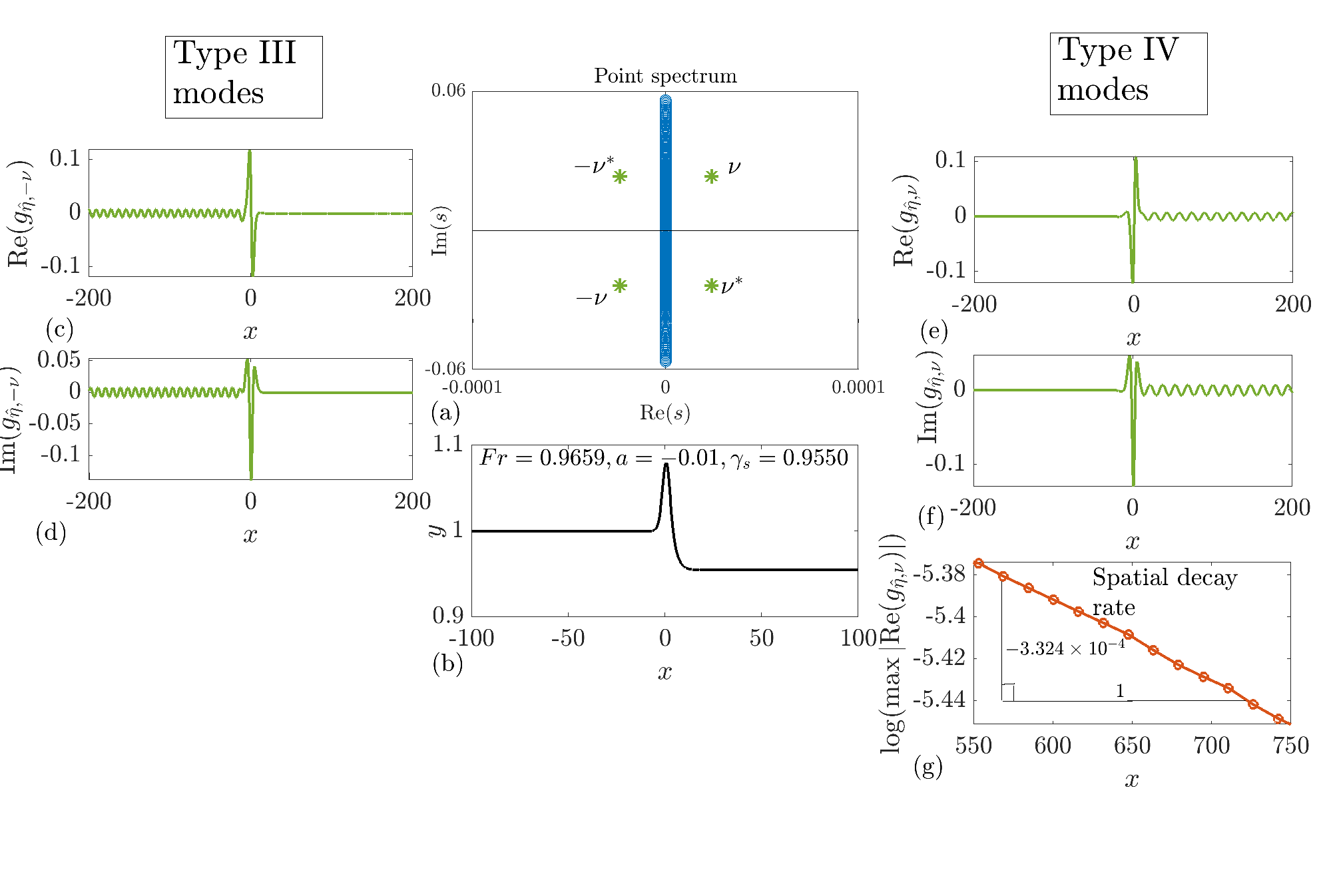}
  \caption{Type \soltype{3}, \soltype{4} modes of the numerically calculated $s_{\mathrm{p}}$ 
 for the hydraulic-fall solution with $Fr=0.9659$, $a=-0.01$, $b=0.3$. Panel (a): $s_{\mathrm{ess}}$ is 
 marked on the imaginary axis and $s_{\mathrm{p}}$ with $\nu = 
 2.375\times 10^{-5} + 0.0235\mathrm{i}$ is indicated with green stars. 
 Panel (b): The underlying steady state. Panels (c)-(d): Type \soltype{3} eigenmodes corresponding to $s_{\mathrm{p}}=-\nu,-\nu^*$ (real and 
 imaginary parts). Panels (e)-(f): Type \soltype{4} eigenmodes corresponding to 
 $s_{\mathrm{p}}=\nu,\nu^*$. Panel (g) $\log(\max|\mathrm{Re}(g_{\hat{\eta},\nu})|)$ plotted against $x$ indicating the small spatial decay rate (estimated in the panel) of the downstream wave in 
 panel (e).}
  \label{fig:spectrum}
\end{figure}

In figure~\ref{fig:spectrum} we show results for the negative forcing case $a=-0.01$, $Fr=0.9659$. In panel (a) the point spectrum
$s_{\mathrm{p}}$ (green stars) is shown together with the essential spectrum $s_{\mathrm{ess}}$ (blue circles) in the complex $s$-plane.  
The underlying steady state is shown in panel (b). Panels (c) and (d) show the corresponding \soltype{3} modes, and panels (e) and (f) show the 
corresponding \soltype{4} modes. Note that the sponge layer is not included for either the type \soltype{3} or the type \soltype{4} calculations.
The exponential decay rate of the type \soltype{4} downstream waves is scrutinised in panel (g) by plotting the logarithm of the local wave 
maxima against $x$ and then estimating its slope. The decay rate is computed to be $-3.3\times 10^{-4}$.
 
\begin{figure}
  \centering
  \includegraphics[width=\textwidth]{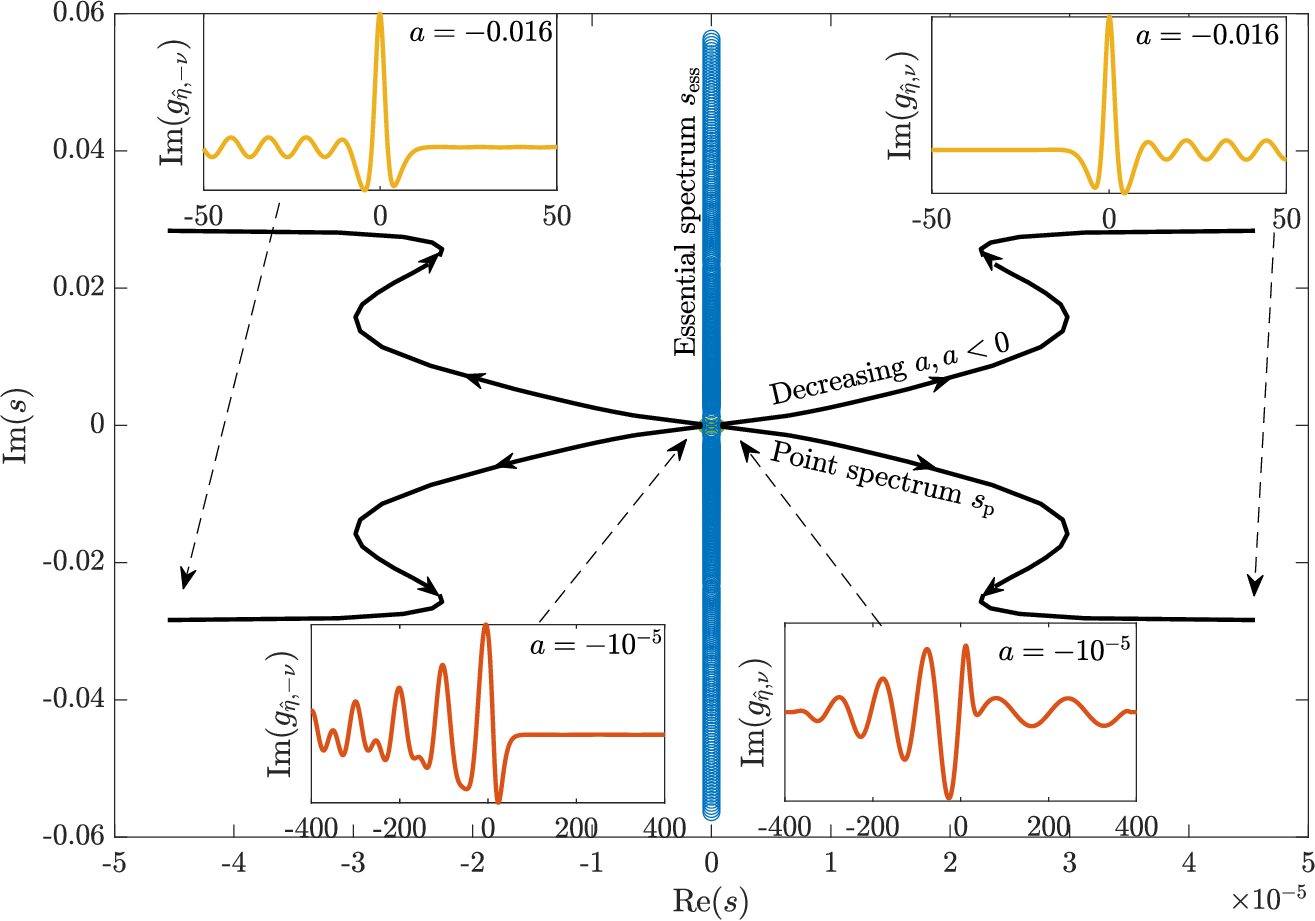}
  \caption{Hydraulic-fall solutions for $b=0.3$. The numerically calculated $s_{\mathrm{p}}$ as $a$ is varied  (the corresponding values of $Fr$ can be inferred from figure~\ref{fig:fnl_hfall}). 
  The solid black lines indicate $s_{\mathrm{p}}$ and the arrows indicate how $s_{\mathrm{p}}$ evolves as 
  $a$ decreases from zero. $s_{\mathrm{ess}}$ is shown by hollow circular markers. We note that the eigenmodes of $s_{\mathrm{ess}}$ 
  change as $a$ is varied but $s_{\mathrm{ess}}$ itself remains on the imaginary axis. The inset diagrams show eigenmodes corresponding to 
  $s_{\mathrm{p}}$ as indicated by arrows for the values of $a = -10^{-5}$ (bottom inset panels) and $a=-0.016\approx a^*$ (top inset panels).}
  \label{fig:vary_alpha_spectrum}
\end{figure}

We only find a non-empty point spectrum when $a<0$ (we calculated the essential and point spectrum for values of $a,b$ in the range $a^*<a<0.1$ and $0.3\leq b\leq 3.0$) and, moreover, that the point spectrum contains only four  
eigenvalues. We can investigate the linear instability for $a<0$ further by computing the point spectrum as $a$ is varied continuously. 
Figure~\ref{fig:vary_alpha_spectrum} 
shows that as $a\to 0^-$ the quartet of eigenvalues appears to approach the origin. This is consistent with the fact that when $a=0$, i.e. for a flat 
bottom, the spectrum contains an isolated zero of multiplicity four \citep[see, for example][]{akers2012spectral}. As $a\to a^*\approx -0.016$, 
the unstable eigenmodes approach those found at the `termination point' solutions discussed in \S~\ref{sec:fnl}. The fact that we 
have not found a point spectrum when $a>0$ appears to suggest that hydraulic-fall solutions over positive bumps are linearly stable. %\mgb{ I'm not sure what to make of the inset eigenmode profiles for the small $a$ values. Can we compare with the literature for $a=0$ perhaps?} 

Although we have analysed the stability properties of the linearised system, as is well known linear stability does not imply nonlinear stability 
\citep[see, for example][]{holm1985nonlinear}. In the next section we shall perform numerical simulations to provide evidence for nonlinear 
stability.
%%%%%%
%\begin{figure}
%  \includegraphics[width=\textwidth]{./figures/convergence_eigenmode.eps}
%  \caption{Hydraulic-fall solution. The numerically calculated eigenmode for $s_{\mathrm{p}}=\nu$ for $a=-0.0001$ for domain lengths $L = 500,1000,2000$.}
%  \label{fig:L_convergence}
%\end{figure}
%%%%%%
%%%%%%%%%%%
%%%%%%%%%%%

%%%%%%%%%%%
%%%%%%
 
%\subsubsection{The linearised solution, $s_{\mathrm{p}}$}
%
%Finally, the general solution to \eqref{pert_equation}, as shown in \cite{changbook} can be written in the form 
%\bea
%\bar{\zeta} = \sum_{k=0}^{\infty}A_k\textbf{g}_{s_{k}}\mbox{e}^{s_{k}t} + \int_{\Lambda}a(s)\textbf{g}_{s}\mbox{e}^{st}\,\mbox{d}s,
%\label{linear_sol}
%\eea
%where the sum is over the eigenvalues in the point spectra ($s_{\mathrm{p}} = s_k$), the region of integration, $\Lambda$, of the integral is over the imaginary axis and $a_k,a(s)$ are expansion coefficients.
%
%By solving the linearised problem we are able to determine the linear stability properties of the hydraulic-fall solutions. However, linear stability does not imply nonlinear stability \citep[see, for example][]{holm1985nonlinear} and therefore in the next section we perform numerical simulations to provide evidence for nonlinear stability.

\subsection{Nonlinear Stability}\label{sec:nonlinear_stability}

In this section we solve an initial value problem in order to confirm the results of the previous section and to probe the nonlinear stability properties of the hydraulic-fall solutions. We introduce a localised perturbation to the steady solution, $y_s(x)$, taking
\begin{equation}
y_f(x,t=0) = y_s(x) + \varepsilon x\mbox{e}^{-a_2^2(x-a_3)^2}.
\label{pert_IVP}
\end{equation}
for some choice of the perturbation amplitude $\varepsilon$ and the constants $a_2$ and $a_3$. 
In the results to be presented below we choose $a_2=0.5$ in \eqref{sig} and set $a_3 = 0$ but the general conclusion on the solution's stability
remains the same for other values of $a_3$.
The computation is carried out as follows. Having deformed the flow domain to incorporate the perturbation \eqref{pert_IVP}, we solve Laplace's equation over this domain together with the steady kinematic condition on the free surface $\Gamma_2$, \eqref{kin_eq}, the no-penetration condition on the bottom $\Gamma_0$, \eqref{bottom_eq}, and the inflow and outflow conditions on $\Gamma_3$ and $\Gamma_1$, \eqref{in_num} and \eqref{out_num}, respectively. This provides us with the initial velocity potential over the perturbed domain, namely $\phi(\textbf{x},t=0)$. We set the downstream target level for the sponge layer as $y_{+} = \gamma_s$.
%We choose a dipole perturbation of the form described in \eqref{pert_IVP} as the perturbation (in an infinite horizontal domain and in the absence of a sponge-layer) contributes zero to the overall mass, given as the integral of $y_f$ over the computational domain. This is important as it easily shown \citep[see, for example][]{benjamin1982,longuet1983} that the mass is a conserved quantity of the problem and hence any candidate perturbation function should preserve this property.

%Second, we perturb the steady solution by a linear combination of eigenmodes extracted from the essential spectrum or the point spectrum, taking
%\bea
%y_f(x,t=0) = y_s(x) + \sum_{\magenta{i=1}}^{N}\varepsilon_ig_{\hat{\eta},s_i}(x),
%\label{pert_eigen}
%\eea
%where $N$ is the number of different modes and the $\varepsilon_i$ are specified amplitudes. This allows us to test more general, non-localised 
%perturbations. We note that if modes from the essential spectrum are included, neither of the initial conditions \eqref{pert_IVP} and 
%\eqref{pert_eigen} satisfy either the inflow condition \eqref{in_num} or the inflow condition \eqref{out_num}. In numerical practice the sponge 
%layers discussed in section~\ref{sec:unsteady_flux} quickly smooth out the upstream waves far upstream and downstream during the first few 
%time-steps.

Our stability analysis of the hydraulic-fall solutions depends on two parameters: the amplitude of the topographic forcing, $a$, and the size of 
the perturbation, $\varepsilon$, in \eqref{pert_IVP}. To standardise the size of the perturbation as $a$ is varied we set
\bea \label{carrot}
\varepsilon = l\Delta_s,\qquad \Delta_s \equiv |1-\gamma_s|,
\eea
with $l>0$. This provides a convenient way to compare results for different values of $a$. %
%\past{For initial conditions in \eqref{pert_eigen} we set the values of $\varepsilon_i$ and use the normalisation that oomph-lib provides: JSK: this needs to be sorted out.}

As the system is Hamiltonian the underlying dynamical system cannot contain attractors or repellors as the Hamiltonian is invariant along any trajectory in the system's phase space. Therefore, in this system, the stability properties of the steady states are interpreted by a local argument; the system can evolve towards/away from a steady state as energy is emitted far upstream/downstream.

%\past{With this in mind we refer tio 

%To probe the nonlinear stability properties, we make use of the concepts of \textit{Lyapunov stability} and \textit{asymptotic stability} \citep[see, for example][p.17]{kuznetsov2013elements}, both defined with reference to a system norm, defined as $d(t)$. A steady state is  \textit{Lyapunov stable}  if $ \forall \epsilon > 0,\:\exists\,\delta>0 \mbox{ such that if } d(t=0) < \delta$, then $\forall t \geq 0,\:d(t) < \epsilon$. In this case the system does not have to settle back to the original steady state but it stays inside a closed ball around it. The stronger notion of \textit{asymptotic stability} means that $\exists\:\delta\:\mbox{such that if } d(t=0)< \delta,$ then $d(t)\to 0\mbox{ as }t\to\infty$. In this case the system is Lyapunov stable but it necessarily returns to the steady state in infinite time.

%Numerically, we can accumulate evidence for either Lyapunov or asymptotic stability by carrying out a large number of calculations for different values of $l$.
When discussing the results of our numerical simulations, it will be useful to refer to the value of the `wave resistance coefficient' that was introduced by \cite{camassa} for the fKdV equation, and which is defined to be
\bea
C_r = -\int_{\Gamma_2}\pdiff{y_{f}(x,t)}{x}y_b(x)\,\mbox{d}S.
\label{wave_coeff}
\eea
In the fKdV model $C_r$ is the power supplied by the bottom forcing, $y_b$ \citep{camassa}. 
%\\[0.1in] \mgb{Presumably asymptotic stability does not apply to a Hamiltonian system. So it feels like there needs to be coherent linkage to the 
%`local argument' in the previous paragraph.}

%%%%%%%%%%%%%
\subsubsection{Positive forcing}

%%%%%%%%%%%%
\begin{figure}
  \centering
  \includegraphics[width=1.0\textwidth]{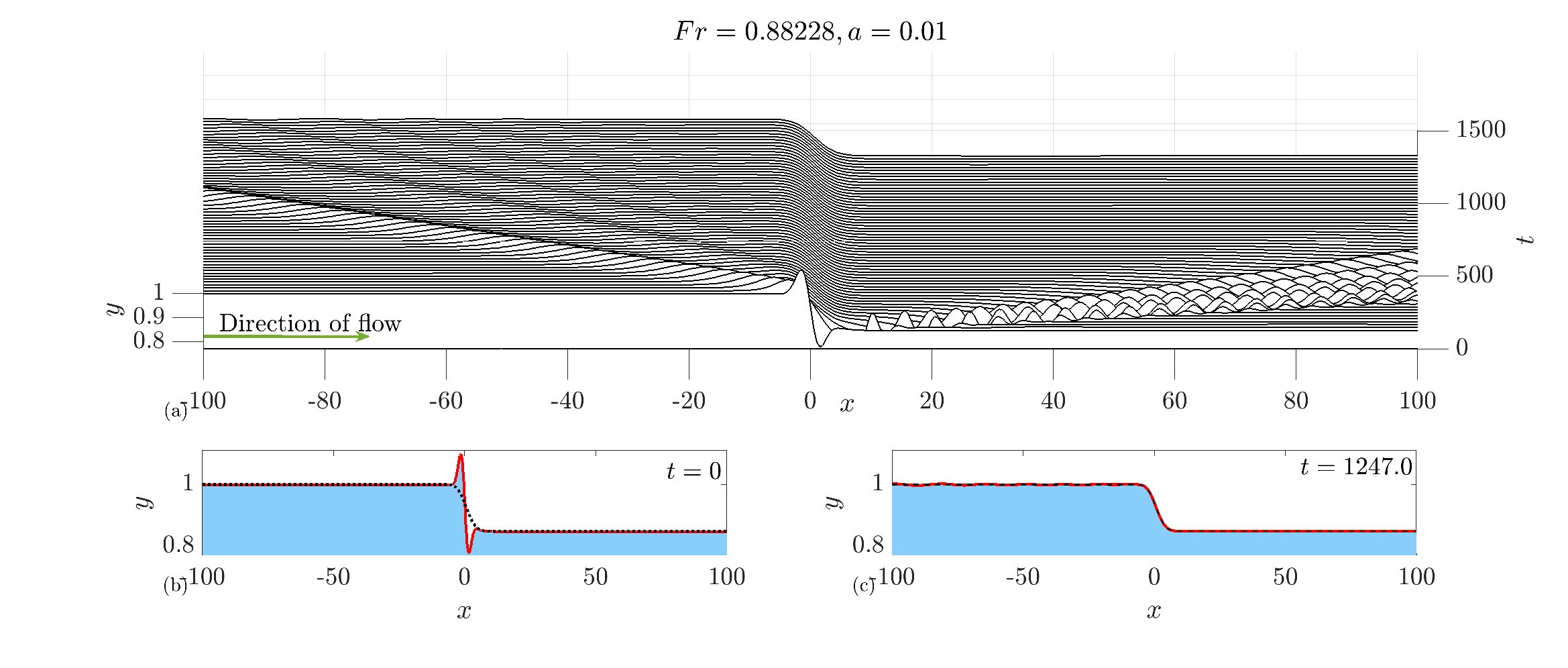}
  \caption{ Perturbation of a hydraulic fall over positive forcing with $a = 0.01$ and $Fr = 0.88228$ and with initial condition \eqref{pert_IVP} with 
  $l = 2.0$, $b=0.3$, $a_2 = 0.5$, and $a_3=0$. (a) Waterfall plot of free-surface evolution. (b) The initial perturbation (solid red line) and the steady state (dotted 
  black line). (c) The free-surface (solid red line) at $t=1247$ and the steady state (dotted black line).}
  \label{fig:alp_0_01_hfall_ic1}
\end{figure}
%%%%%%%%%%%%
We start by examining hydraulic falls over a positive forcing with $a>0$. Figure~\ref{fig:alp_0_01_hfall_ic1} shows the results of a 
time-dependent calculation using the initial condition \eqref{pert_IVP} with $l=2.0$ (other parameter values are quoted in the caption).
An animation of the time evolution of the free surface can be found in the supplementary material (\texttt{movie\_1.mp4}). Evidently the system quickly returns to the original unperturbed stable state. It was found that this behaviour was typical for forcing amplitudes in the range $a=(0, 0.1]$ and perturbations in the range $l \in (0,4]$, providing evidence that the solution is stable to localised perturbations as prescribed in \eqref{pert_IVP} for all $a>0$ sampled.

\subsubsection{Negative Forcing}

\begin{figure}
  \centering
  \includegraphics[width=\textwidth]{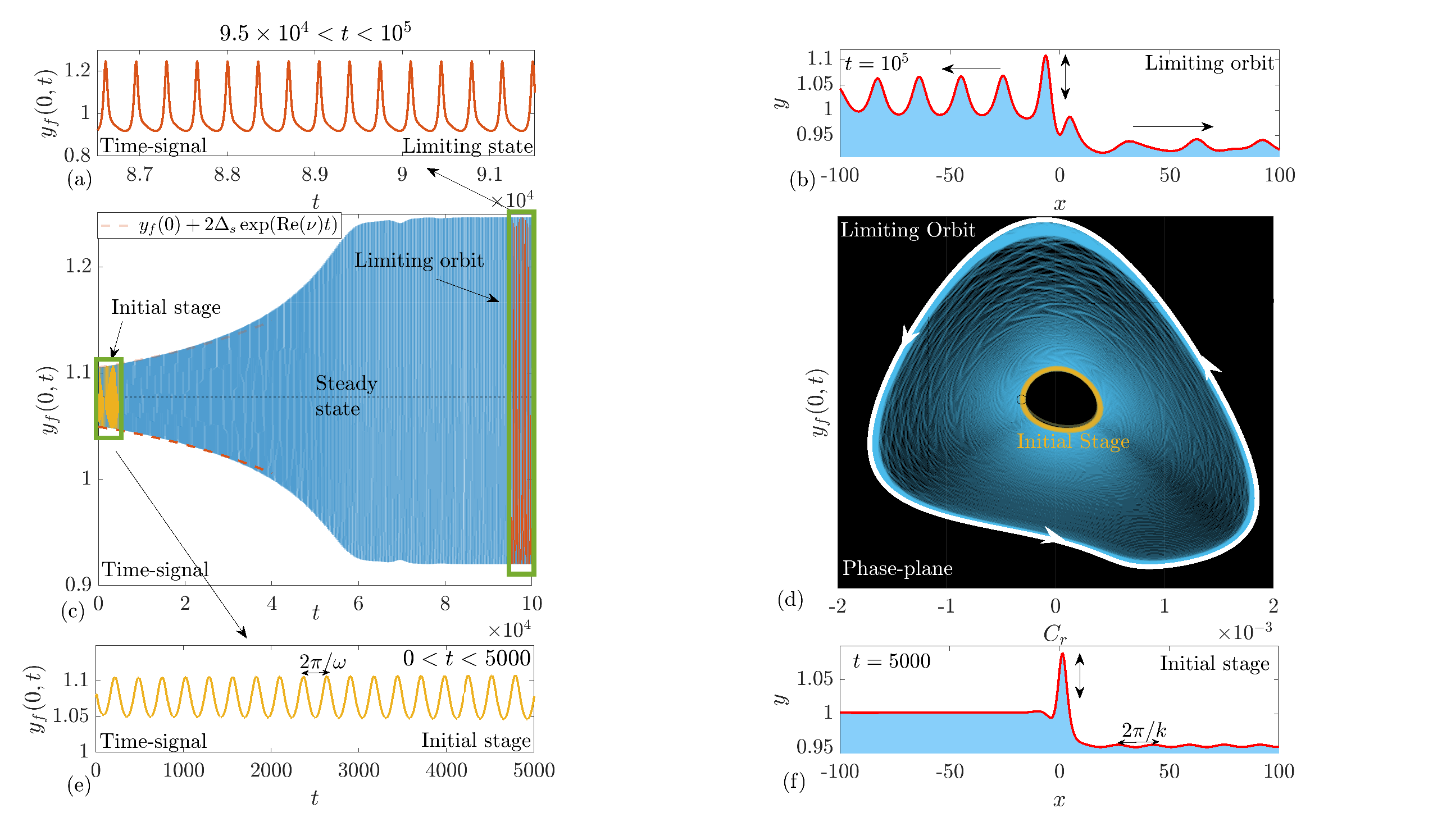}
  \includegraphics[width=\textwidth]{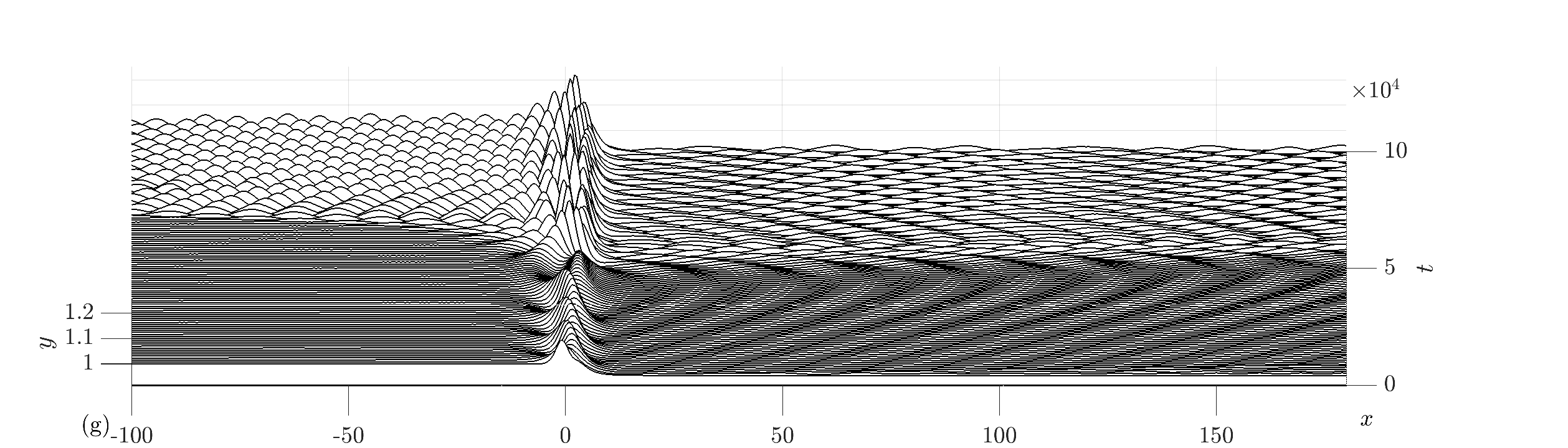}
  \caption{Perturbation of the hydraulic fall solution with $a = -0.01$, $Fr = 0.9659$ (and $\Delta_s = 0.0450$) with the initial condition \eqref{pert_IVP} taking $l = 2.0$, $b=0.3$, $a_2=0.5$, and $a_3=0$. The underlying type \soltype{4} eigenmode of the steady state has $\nu = 2.375\times 10^{-5} + 0.0235\mathrm{i}$. (a) Time trace of $y_f(0,t)$ over the window $8.6\times 10^4<t<9.15\times 10^5$. (b) Free-surface profile at $t=10^5$. The estimation of the wave number, $k$, is indicated. (c) Time trace of $y_f(0,t)$ over $0<t<10^5$. The dotted line indicates $y_{s,f}(0)$. We note the frequency of the oscillations cannot be resolved on this time-scale and zoomed regions are shown in panels (a) and (e). (d) The phase plane projection $(C_r,y_f(0,t))$ with the vertical axis on the same scale as panel (c). Thicker lines indicate locations in phase space more frequently visited. (e) Time trace of $y_f(0,t)$ over $0<t<5000$. (f) Free-surface profile at $t=5000$. The arrows indicate the direction of wave propagation. (g) Waterfall plot of free-surface evolution.  }
  \label{fig:alp_n0_01_hfall}
\end{figure}

Figure~\ref{fig:alp_n0_01_hfall} shows the result of a time-dependent calculation for a topographical dip with $a=-0.01$, $b=0.3$ and initial condition \eqref{pert_IVP} (the other parameter 
values are shown in the captions). We strongly recommend the reader to watch \texttt{movie\_2.mp4} in the supplementary material to support 
the following discussion. A waterfall plot of the time-evolution of the free surface is shown in panel (g). Panel (c) shows the time-trace of $y_f(0,t)$ over the range $0<t<10^5$ (zoomed-in regions of this time-trace are shown in panels (a) and (e)), and panel (d) shows the $(C_r,y_f(0,t))$ phase plane projection. The thicker lines in panel (d) indicate regions in phase space that are visited more frequently. Wave profiles at times $t=5000$ and $t=10^{5}$ are shown in panels (f),(b), respectively.
%%%%%%%
%%%%%%%

\begin{figure}
  \includegraphics[width=\textwidth]{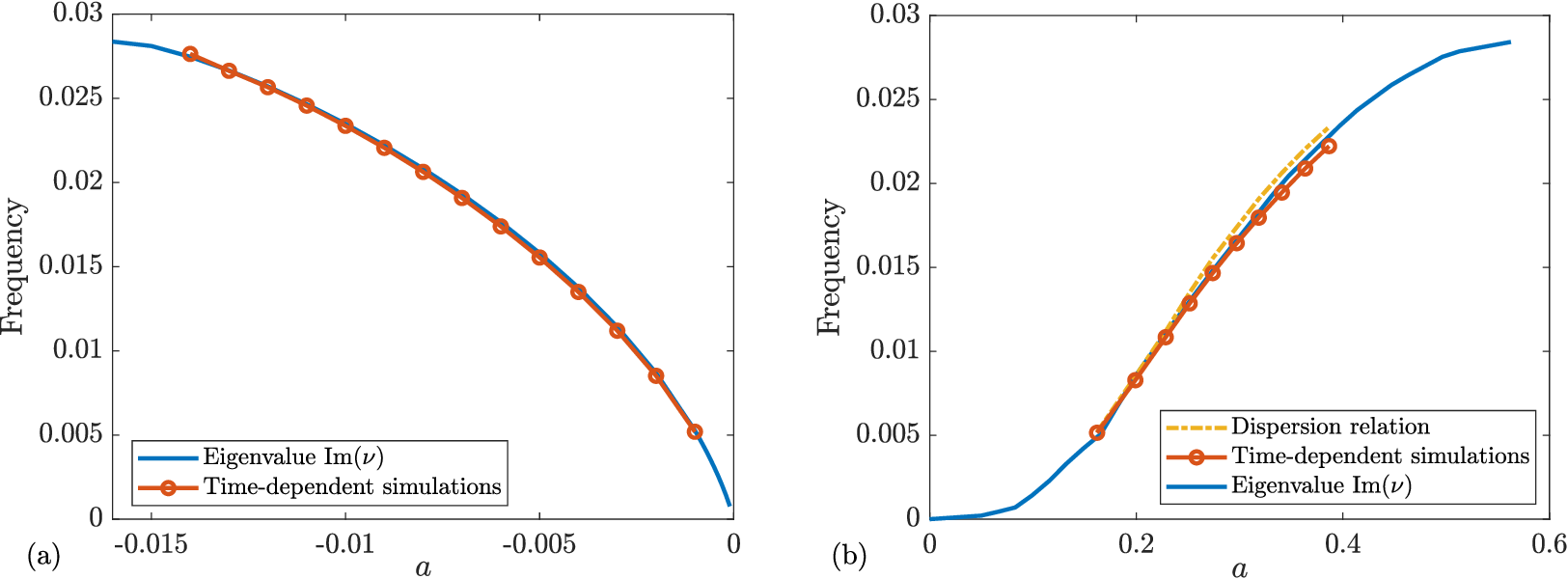}
\caption{(a) The frequency of oscillations in the initial stage of the evolution for the case studied in figure~\ref{fig:alp_n0_01_hfall} but over a range of $a$ values, $b=0.3$. The solid lines are the imaginary part of $\nu$ and the line with circular markers is the estimate of the frequency from the time-dependent simulations. (b) The linear dispersion relation $\omega=\omega(k)$, with $k=k_d$, according to \eqref{downstream_disp}, shown with a dot-dashed curve. The estimated frequency-wave number relation for the small amplitude waves emitted in panel (f) of figure~\ref{fig:alp_n0_01_hfall} is illustrated with red circular markers. The (positive) imaginary part of the eigenvalue in the point spectrum, 
$\mbox{Im}(\nu)$, is shown with a solid blue line.}
    \label{fig:frequency_comparison}
\end{figure}

Considering these results, the time-evolution can be broadly divided into two distinct stages. The initial stage (over $0<t<5000$, say, and shown in panel (e)) is characterised by a roughly sinusoidal temporal oscillation of $y_f(0,t)$ about the steady state value, the latter shown as a dotted line in panel (c). In this stage, the transient growth is well matched by the real part of the unstable eigenvalue, $s=\nu$, of the underlying state as shown by the dashed line in panel (c). The frequency of the oscillation is in excellent agreement with the imaginary part of $\nu$. In fact, the same observation holds for a broad range of topography amplitudes $a$. This is illustrated in figure~\ref{fig:frequency_comparison}(a), where $\mbox{Im}(\nu)$ is plotted against $a$. Also shown is the frequency of $y_f(0,t)$ estimated from the time-trace during the initial stage of the evolution. 
Evidently there is excellent agreement between the two. We also found that this frequency is independent of the perturbation amplitude $l$. %\past{A final observation is that this frequency and the value of $|\mathrm{Im}(\nu)|$ is always larger in magnitude to the absolute value of $\tau$, as shown by the dotted line in figure~\ref{fig:frequency_comparison}(a).}
%%%%%%%%%%%%%%%%%%
%%%%%%%%%%%%%%%%%%

During this initial stage of the evolution, as the amplitude of the oscillations increases and small amplitude waves are emitted and travel 
downstream, as can be seen in panel (f) of figure~\ref{fig:alp_n0_01_hfall}. We would expect that the wave number of these emitted waves
conforms with the linear dispersion relation \eqref{downstream_disp}. 
%, with wave number $k$, which we 
We can test this by estimating the frequency and wave number of the emitted waves from the time signals and from the instantaneous 
free-surface profile, respectively. To approximate the frequency we calculated the mean separation of the local maxima of the time-signal 
$y_f(0,t)$, and to estimate the wave number we extracted the mean separation of the local maxima in the spatial wave profile $y_f(x,t)$ at 
$t=10^4$. The results are shown in figure~\ref{fig:frequency_comparison}(b), where we see good agreement with the dispersion relation  $
\omega=\omega(k)$, with $k=k_d$, in \eqref{downstream_disp}, although the disparity between the two increases as the wave number $k$ 
increases. Also shown in the figure~\ref{fig:frequency_comparison}(b) is the (positive) imaginary part of the eigenvalue in the corresponding point 
spectrum (see section~\ref{sec:pointspec}).

Eventually the amplitude of the oscillations saturate heralding the second evolutionary stage, which we refer to as the limiting stage. Here the 
oscillations are less sinusoidal in character. The limiting stage is characterised by a regular stream of larger-amplitude cnoidal waves propagating in the upstream direction and away from the topographical dip, 
and, in contrast to the initial stage, the waves travelling downstream, as shown in panel (b) of figure~\ref{fig:alp_n0_01_hfall}, now consist of 
multi-harmonic waves. The phase-plane projection $(C_r,y_f(0,t))$ shown in panel (d) of the 
same figure has a distinctive guitar-pick shape. At large time $t$ the trajectory appears to have converged to a closed loop (shown in white with 
arrows), which represents an invariant solution of the system. We remark that this limiting behaviour was found to be typical for parameter values $a^*<a<0$, $0.3\leq b \leq 3.0$. We also recommend the reader to watch the animation in \texttt{movie\_3.mp4} focussing on how the limiting stage for figure~\ref{fig:alp_n0_01_hfall} evolves over a temporal period.

%If instead of \eqref{pert_IVP} the eigenmode initial condition \eqref{pert_eigen} is used then broadly similar results are observed. In particular, if the unstable eigenmode $g_{\hat{\eta},\nu}$ is used as the perturbation in \eqref{pert_eigen}, the system eventually settles down to the same invariant solution seen in figure~\ref{fig:alp_n0_01_hfall}. If the stable eigenmode, $g_{\hat{\eta},-\nu}(x)$, is used, the upstream waves in the eigenmode profile (\magenta{see panel (e) of figure~\ref{fig:spectrum}}) quickly decay in amplitude and the system settles \magenta{to the same invariant solution at large time.}
In conclusion the steady state hydraulic-fall solution with negative forcing is unstable and perturbations from it settle down to a time-dependent, nonlinear stable invariant solution irrespective of size of the perturbation in \eqref{pert_IVP}. We discuss this in more detail in \S~\ref{sec:unsteady_inverse}.

\subsubsection{Hydraulic-rises}

\begin{figure}
  \includegraphics[width=\textwidth]{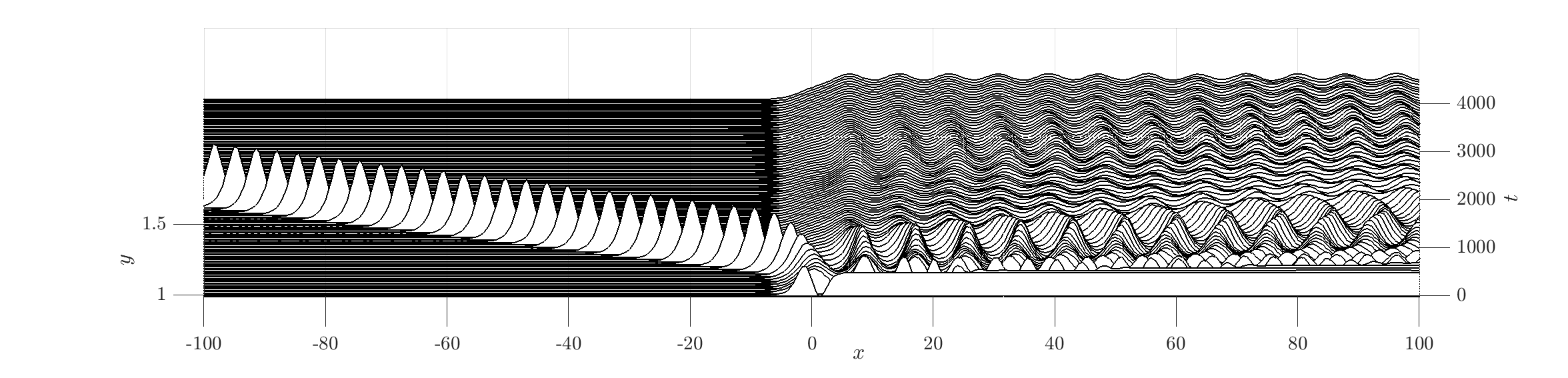}
  \caption{Waterfall plot of free-surface evolution of a perturbation of a hydraulic rise over positive forcing with $a = 0.01$ and $Fr = 1.1173$ and with initial condition \eqref{pert_IVP} with 
  $l = 2.0$, $b=0.3$, $a_2 = 0.5$, and $a_3=0$.}
  \label{fig:hrise_pos}
\end{figure}

\begin{figure}
  \includegraphics[width=\textwidth]{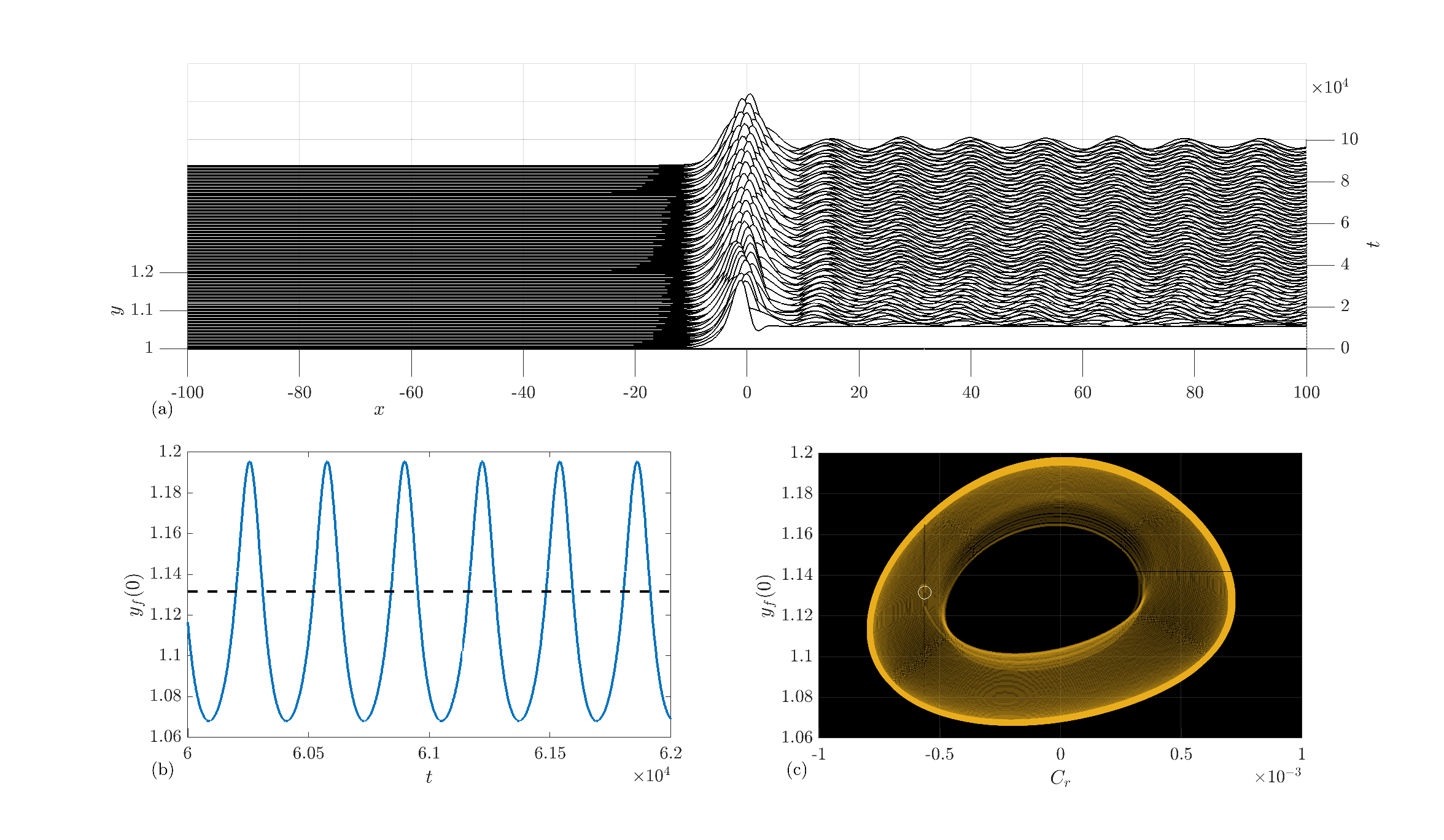}
  \caption{Hydraulic rise. The case $a=-0.01$, $b=0.3$, $Fr = 1.0436$ and with initial condition \eqref{pert_IVP} with $l = 2.0$, $a_2 = 0.5$, and $a_3=0$. (a) Waterfall plot for the system starting from a state of near uniform flow.  (b) Time-trace of $y_f(0,t)$ in the limiting stage. (c) The projected phase plane, $(C_r,y_f(0,t))$. The initial condition is shown as a circular marker, and the trajectory is in orange.}
  \label{fig:hrise_neg}
\end{figure}

For completeness, we briefly describe the nonlinear stability of the steady hydraulic-rise solutions. Hydraulic rises are unstable for both positive 
and negative $a$ as demonstrated in figures~\ref{fig:hrise_pos} and \ref{fig:hrise_neg}, respectively. For positive forcings, as the initial 
disturbance propagates downstream a solitary-wave emanates from the origin and moves upstream whilst a steady cnoidal-wave pattern 
emerges downstream. Ignoring the solitary wave, this structure is known as a generalised steady hydraulic rise (see figure~\ref{fig:hrise_pos}). 
For negative forcings, similar to what was found for the hydraulic fall, the system settles into a time-periodic orbit as indicated in panel (c) of figure~\ref{fig:hrise_neg}. We observe that the temporal signal of 
$y_f(0)$ in the limiting stage, as shown in panel (b), is more sinusoidal than that for the hydraulic fall (see panel (a) of figure~\ref{fig:alp_n0_01_hfall}), and that a similar cnoidal wave pattern to that seen for the positively forced hydraulic rise emerges downstream.

%\mgb{Is the system not Lyapunov stable by definition as it is energy conserving?}
%\\[0.1in]

%%%%%%%%%%
\subsection{Time-dependent invariant solution}\label{sec:unsteady_inverse}

%%%%%%%%%%%%%%%%%%
\begin{figure}
  \includegraphics[width=\textwidth]{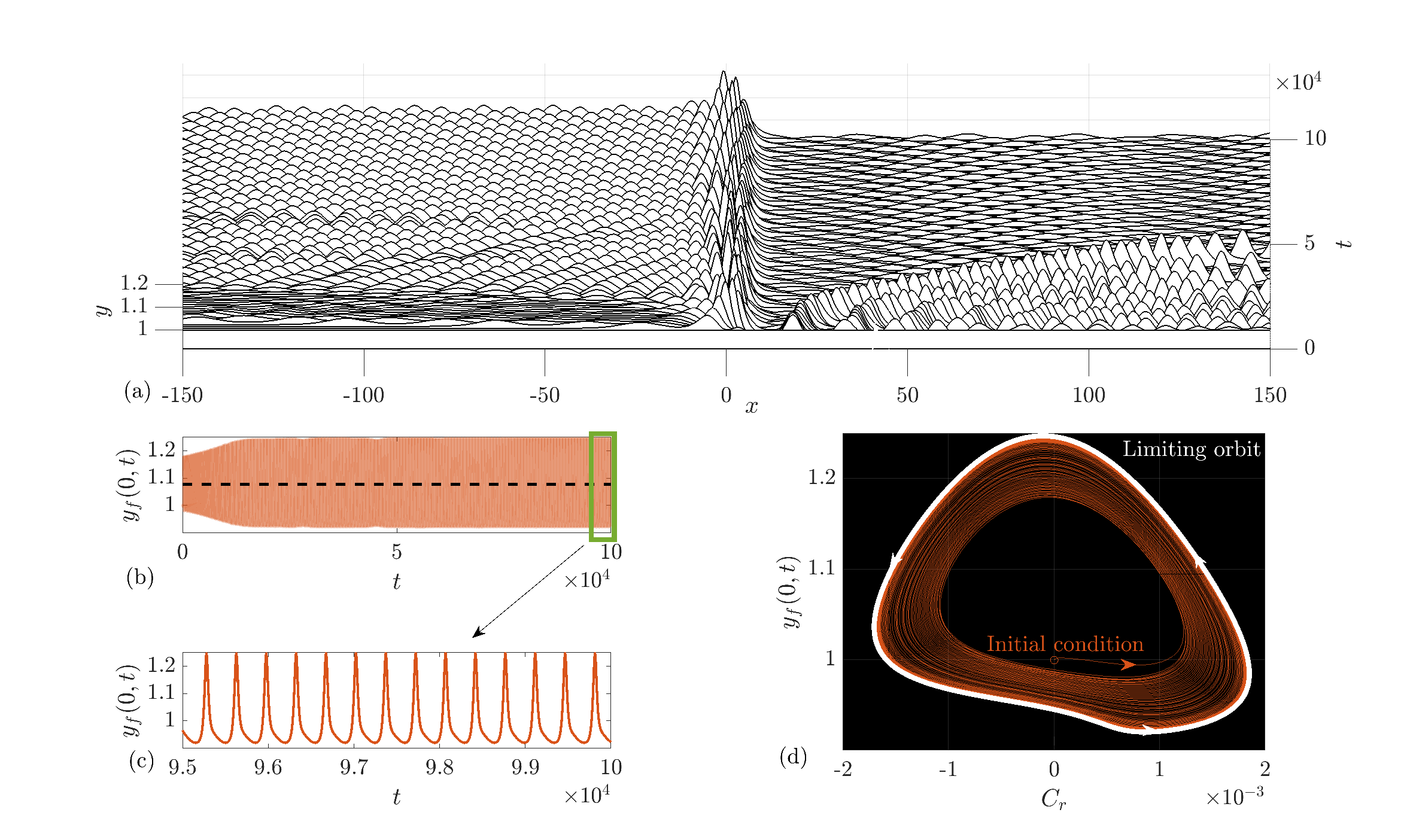}
  \caption{Hydraulic fall. The case $a=-0.01$, $b=0.3$, $Fr = 0.9659$. (a) Waterfall plot for the system starting from a state of near uniform flow.  (b) Time-trace of $y_f(0,t)$. (c) Time-trace in the period $9.5\times 10^{4}<t<10^{5}$. (d) The projected phase plane, $(C_r,y_f(0,t))$. The initial condition is shown as a circular marker, the system trajectory is in red, and the limiting orbit from figure~\ref{fig:alp_n0_01_hfall} is shown in white.}
  \label{fig:alpha_neg_flat}
\end{figure}  
%%%%%%%%%%%%%%%%%%

To further explore the existence of a time-dependent invariant solution for negative forcing, $a<0$, we calculate the free-surface response when 
the system starts from a condition of near uniform flow with a flat free surface. Specifically, to provide the initial condition we solve Laplace's equation over the flow domain shown in figure~\ref{fig:problem_domain} but with a flat free surface $\Gamma_2$, enforcing
the steady form of the free-surface kinematic condition, \eqref{kin_eq}, the no-penetration condition on the bottom $\Gamma_0$, \eqref{bottom_eq}, and the inflow and outflow conditions on $\Gamma_3$ and $\Gamma_1$, \eqref{in_num} and \eqref{out_num}, respectively. This yields the initial velocity potential $\phi(\textbf{x},t=0)$. This starting condition is artificial as the pressure at the free surface does not match the constant ambient pressure at $t=0$, but we overlook this slight inconsistency. During the calculation we set the downstream sponge target level as $y_{+} = 1$.
%To further explore the existence of a time-dependent invariant solution for negative forcing, $a<0$, we calculate the free-surface response when 
%the system starts from a condition of \magenta{near} uniform flow with a flat free surface. We achieve this by deforming the flow domain 
%with the topographic forcing and flat free-surface and then solve Laplace's equation over this domain together with the steady kinematic condition on the free surface $\Gamma_2$, \eqref{kin_eq}, the no-penetration condition on the bottom $\Gamma_0$, \eqref{bottom_eq}, and the inflow and outflow conditions on $\Gamma_3$ and \magenta{$\Gamma_1$}, \eqref{in_num} and \eqref{out_num}, respectively. This provides us with the initial velocity potential over the perturbed domain, namely $\phi(\textbf{x},t=0)$. In this calculation we set the downstream sponge target level as $y_{+} = 1$.

The results are shown in 
figure \ref{fig:alpha_neg_flat} for $a=-0.01$ and $Fr=0.9659$, which are the same values as in 
figure~\ref{fig:alp_n0_01_hfall}. The waterfall plot shown in panel (a) indicates that the region left in the wake of the downstream travelling waves is itself wavy. The signal $y_f(0,t)$ shown in panels (b),(c) of figure~\ref{fig:alpha_neg_flat} appears 
to be periodic and it is similar in character to the oscillations observed in figure~\ref{fig:alp_n0_01_hfall}(a). Furthermore, in panel (d) of 
figure~\ref{fig:alpha_neg_flat}, the trajectory in the $(C_r,y_f(0,t))$ plane is shown and it is clear that the system is evolving towards the 
guitar-pick shaped limiting orbit observed in figure~\ref{fig:alp_n0_01_hfall}, which is also shown in the panel (coloured white and labelled as the 
`limiting orbit'). This calculation provides further evidence of what appears to be a time-periodic orbit and that it is the same solution as in 
figure~\ref{fig:alp_n0_01_hfall}.

\begin{figure}
  \centering
  \includegraphics[width=\textwidth]{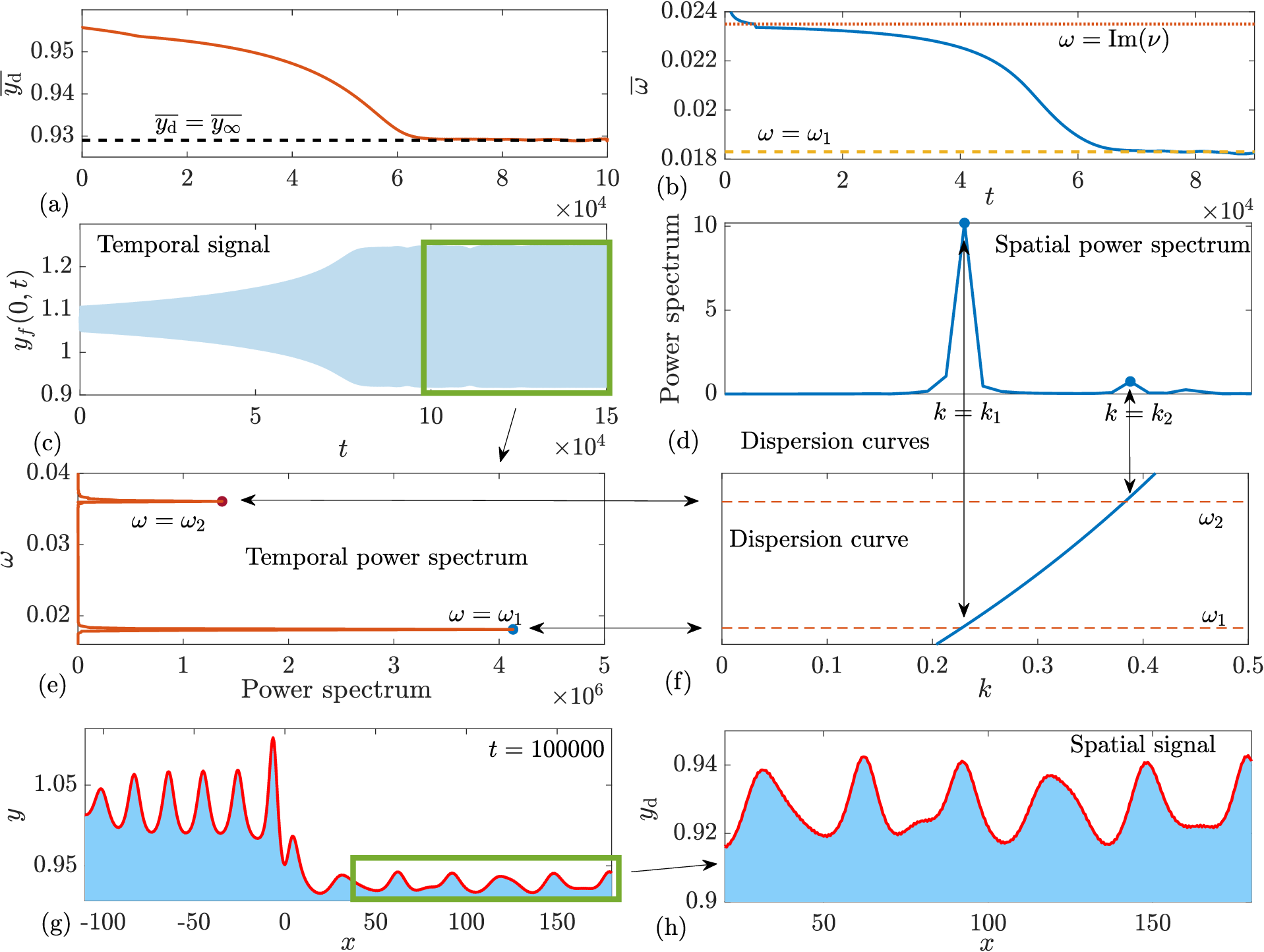}
  \caption{Frequency and wave number analysis of the downstream portion of the wave, $y_{\mathrm{d}}$, and the time-signal $y_f(0,t)$, for the calculation in figure~\ref{fig:alp_n0_01_hfall}. Panel (a): the downstream mean level, $\overline{y_{\mathrm{d}}}$ as a function of time. Panel (b) The average frequency as $t$ varies, estimated from the local maxima of the time signal of $y_f(0,t)$. Panel (c): the time signal of $y_f(0,t)$. Panel (d): the power spectrum of the down stream wave in the limiting stage of the evolution with the two peaks occurring at $k_1 \approx 0.2275,k_2\approx 0.3850$. Panel (e): the power spectra of $y_f(0,t)$, note the vertical axes is the frequency, $\omega$, and the horizontal axes is the power. Panel (f) The limiting stage dispersion curve together with the levels $\omega = \omega_1\approx 0.018,\omega_{2}\approx 0.036$. Note the horizontal axis and scale are identical in panels (d) and (f) and the vertical axis and scale are identical in panels (e) and (f) to aid comparison. Panels (g) and (h): the wave profile at the given time slot and the downstream portion of the free-surface.}
  \label{fig:limiting}
\end{figure}

In figure~\ref{fig:limiting} we examine the time-dependent invariant solution further by analysing the calculation in figure~\ref{fig:alp_n0_01_hfall}
for $a=-0.01$. We begin by noting that, as can be seen in figure~\ref{fig:limiting}(a), the mean level of the downstream spatial signal, denoted $\overline{y_{\mathrm{d}}}$, decreases as time progresses. Consequently, a modified downstream dispersion relation obtains for the limiting stage, and this is found 
by replacing $\gamma_s$ and $V$ in \eqref{downstream_disp} with $\overline{y_{\infty}}$ and $1/\overline{y_{\infty}}$ respectively.

Next we observe that the average temporal frequency, denoted $\overline{\omega}$, estimated using the local maxima of the time
signal $y_f(0,t)$ in figure~\ref{fig:limiting}(c), varies in time from a value close to $\overline{\omega} \approx \mathrm{Im}(\nu)$ in the initial 
stage of evolution to a smaller value, denoted $\omega_1$, in the limiting stage (see panel (b) of figure~\ref{fig:limiting}). This frequency 
modulation can be explored further by calculating the power spectrum of the temporal signal of $y_f(0,t)$ in the limiting stage, i.e. over 
$10^5<t<1.5\times 10^5$. The result is shown in panel (e) of figure~\ref{fig:limiting}. 
This shows that there are actually two main temporal frequencies in the limiting stage as indicated by the 
peaks in the temporal power spectra shown in panel (e) of figure~\ref{fig:limiting}. We denote these two dominant frequencies by $\omega_1\approx  
0.018$ and $\omega_2\approx  0.036$.

In addition, we find there are two dominant wave numbers of the spatial downstream wave signal, $y_{\mathrm{d}}$ (see panel (h) of  
figure~\ref{fig:limiting}), as indicated by the spatial power spectrum of $y_{\mathrm{d}}$ at $t=10^5$ shown in panel (d) of figure~\ref{fig:limiting}. 
We denote these two dominant wave numbers by $k_1\approx 0.2275$ and $k_2\approx 0.3850$.

Bringing this all altogether, in panel (f) of figure~\ref{fig:limiting} we show the limiting dispersion curve together with two horizontal lines indicating 
the levels $\omega = \omega_1,\omega_2$. The values of $k$ where the curve intersects these lines correspond precisely to the two dominant 
wave numbers $k_1$, $k_2$ seen in panel (d) of figure~\ref{fig:limiting}. To aid comparison across panels in the figure the horizontal axes in 
panel (d) and panel (f) of figure~\ref{fig:limiting} are identical, as are the vertical axes of panel (e) and panel (f). These results indicate that the wave numbers of the downstream disturbance are linked to the temporal frequencies via the downstream dispersion curve.

%These results indicate that 
%the long-term invariant solution is temporally biharmonic \mgb{probably all harmonics are present but with diminishing amplitudes?} 
%and the wave numbers of the downstream disturbance are linked to the temporal frequencies via the downstream dispersion curve.

\begin{figure}
  \centering
  \includegraphics[width=\textwidth]{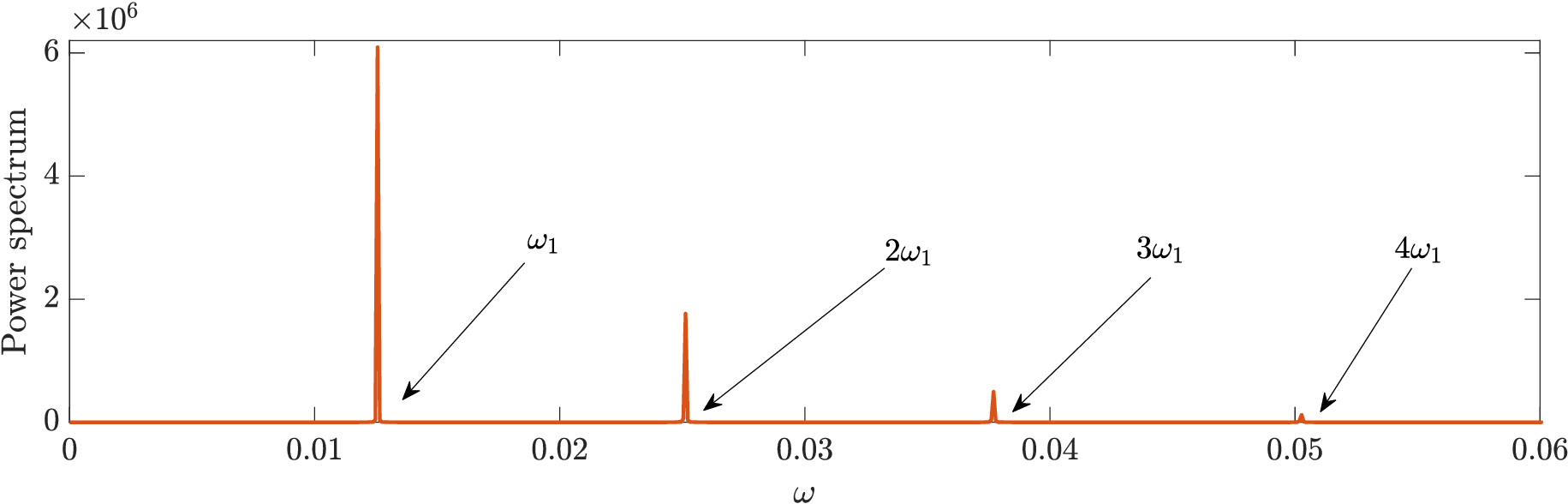}
  \caption{The power spectrum of $y_f(0,t)$ in the limiting stage of the evolution for the solution when $a=-0.005, b=0.3$, $l=2$. The primary harmonic, $\omega_1\approx 0.01257$, and the next four subharmonic frequencies are identified.}
  \label{fig:multi}
\end{figure}

In figure~\ref{fig:multi} we show the power spectrum of the time-signal of $y_f(0,t)$ at the limiting stage for a slightly different case with 
$a=-0.005$, 
$b=0.3$. The energy peak at the dominant frequency $\omega_1$ is evident, with other peaks of diminishing strength located approximately at 
$\omega = n\omega_1$, where $n=2,3,4,\cdots$, presumably generated via nonlinear effects. This is strong numerical evidence that the invariant solution is a periodic orbit. The downstream wave numbers associated with the subharmonic frequencies, $n\omega_1$, must satisfy the linear dispersion relation \eqref{downstream_disp} and, in general, will be incommensurate. This explains the rather irregular appearance of the downstream waves in panel (h) of figure~\ref{fig:limiting}.
%%%%%%%%%%%%%%
%%%%%%%%%%%%

In summary, in this subsection we have obtained the most important result of the present article, namely the apparent existence of a 
temporally periodic orbit of the fully nonlinear system. Time-periodic behaviour in free-surface flow over topography has been 
observed before, notably for the forced KdV equation \cite{camassa}, \cite{wu1987generation} and 
\cite{grimshaw1986}. Through unsteady simulations these authors found a similar oscillatory behaviour of the free-surface over the topography. 
In this case the period of the oscillation was linked to the frequency at which solitary waves were emitted in the upstream direction. However, in none of these studies was the long time behaviour explored.

%\past{JSK: Get rid of sentence:}
%The fKdV model loses validity in the large-time limit \mgb{why if the solution is periodic?} \past{JSK: Yes, good point} but the short-time FNL results presented here are consistent with these studies \mgb{can you say in what way?} . 

\section{Conclusion}\label{sec:discussion}

We have examined the stability of a hydraulic-fall flow over a localised bottom topography. The topography is in the form of a  
gaussian whose amplitude is either positive (a bump) or negative (a dip). In the initial part of our analysis we studied the linear stability of these 
solutions by first demonstrating that the full equations can be cast in the form of a canonical Hamiltonian system with a Hamiltonian 
that is a modification of that derived by \cite{zakharov1968stability} and \cite{craig1993numerical} 
for the classical water-wave problem. As a consequence the spectrum of the system linearised about a steady hydraulic-fall solution has a 
four-fold symmetry in the complex plane. 

In the case of a bump, our numerical computations showed that the point spectrum is either empty or else all of its eigenvalues are located on the imaginary axis so that the underlying flow is 
spectrally stable. For a dip we identified a single eigenvalue in the point spectrum (to within the four-fold symmetry) that lies off the imaginary axis 
with a positive real part, meaning that the flow is linearly unstable. For both the bump and the dip the essential spectrum occupies the whole of 
the imaginary axis. For both the bump and the dip we carefully analysed the associated eigenfunctions and we demonstrated
that their behaviour in the far-field, away from the localised topography, could be reconciled with the dispersion relation for small amplitude waves 
over a flat bottom.

In the second part of our work we examined the nonlinear stability properties by solving an initial value problem for the full Euler equations, with 
the initial condition taken to represent a small perturbation about a steady hydraulic fall. For a bump the simulations showed that the 
disturbance to the 
free surface disperses so that the steady solution is ultimately recaptured. This is in keeping with the prediction of the linear stability analysis. 
Consequently, we concentrated most of our attention on the dip. In this case our numerical 
simulations showed that the steady hydraulic fall is also nonlinearly unstable and that, ultimately, the flow approaches a periodic state in which the free surface exhibits a localised peak that pulsates up-and-down over the dip, while small amplitude waves are emitted continuously 
in both the upstream and the downstream 
directions. This periodic state may be interpreted as an invariant solution of the underlying dynamical system.
We have shown numerically that the periodic waves that are emitted downstream have a dominant temporal frequency that can 
be accurately linked to the linear dispersion relation based on the mean downstream depth. The subharmonic frequencies of these 
waves were found to be rational multiples of the dominant frequency, thereby reinforcing our conclusion that a periodic state has 
been achieved.

%\vspace{0.125in}
%The possibility that the could become quasi-periodic as we vary parameters or add more physical effects cannot be discounted and there is an intriguing possibility that this solution could become an invariant torus of the system; a solution that has not been identified before in the literature for this problem. This solution could also be related to the complicated dynamics observed in \cite{keelercriticalflow} at $Fr=1$ and also for supercritical flow in \cite{camassa,grimshaw1986}. It remains an interesting possibility that in these situations and in other physical regimes, this invariant solution could dominate the dynamics of the system, an intriguing possibility that deserves further attention.

Interesting questions remain. For example, beside periodic solutions as identified here, there exists the possibility of quasiperiodic solutions. Such objects exist in phase space as so-called invariant tori \cite[e.g.][]{kuznetsov2013elements}  . However, 
directly calculating time-dependent invariant solutions, such as periodic-orbits or invariant tori, remains a highly non-trivial numerical task. In other applications in fluid dynamics, periodic orbits have been computed semi-analytically using a weakly nonlinear approach, e.g. for air bubbles in a Hele-Shaw channel \citep{keeler2019invariant}. \cite{page2018koopman,page2019koopman} have recently applied Koopman analysis and dynamic mode decomposition to approximate periodic orbits for Burger's equation and the Navier-Stokes equations. 
It would be of interest to apply such methods to the free-boundary problem considered here in order to further probe the landscape of invariant solutions in phase space. 

Finally, as was noted in the Introduction, there is an apparent lack of experimental results in the literature concerning hydraulic falls over a dip. It 
would be of interest to see if the numerical observations in the present work, and in particular the time-periodic flow that appears at large time, 
can be realised in the laboratory.

%\mgb{Add more on numerics earlier} %The key advantage of the FEM approach is that although there are more unknowns, compared with the boundary-integral method that only calculates quantities on the free-surface, the Jacobian matrix from this approach is sparse; enabling efficient inversion  with state-of-the-art linear solvers such as SuperLu \citep{li2005overview}. Furthermore, although explicit time-stepping has been used in the boundary-integral studies referenced above, the sparsity of the Jacobian matrix means that implicit time-steppers (here we implemented the second-order BDF method) can integrate the system using higher time-steps and the long-term evolution can be accurately and efficiently probed.
%We remark that this FEM formulation can, in principle, be simply extended into the 3D steady and time-dependent water-wave problem. 

%\backsection[Funding]{J. S. K acknowledges funding from the Leverhulme Trust; ECF-2021-017}
%\backsection[Declaration of Interests]{The authors report no conflict of interest.}
%\backsection[Data availability statement]{The code that support the findings of this study are openly available in figshare at {\tt https://doi.org/10.6084/m9.figshare.22128548.v1} \past{\textbf{JSK: MUST DO THIS WHEN READY}}}
%\backsection[Author ORCID]{ J. S. K: {\tt https://orcid.org/0000-0002-8653-7970},\\ M. G. B: {\tt https://orcid.org/0000-0002-7480-4138} }
%\backsection[Author Contributions]{Both authors contributed to developing the theory and results in this paper. The numerical model and numerical calculations were developed and performed by JSK.}
  
\vskip2pc

%\appendix

\bibliographystyle{jfm}
\bibliography{water_wave}

\begin{thebibliography}{44}
\expandafter\ifx\csname natexlab\endcsname\relax\def\natexlab#1{#1}\fi
\def\au#1{#1} \def\ed#1{#1} \def\yr#1{#1}\def\at#1{#1}\def\jt#1{\textit{#1}}
  \def\bt#1{#1}\def\bvol#1{\textbf{#1}} \def\vol#1{#1} \def\pg#1{#1}
  \def\publ#1{#1}\def\arxiv#1{#1}\def\org#1{#1}\def\st#1{\textit{#1}}

\bibitem[Akers \& Nicholls(2012)]{akers2012spectral}
{\sc \au{Akers, B.} \& \au{Nicholls, D.~P.}} \yr{2012}  \at{Spectral stability
  of deep two-dimensional gravity water waves: repeated eigenvalues}.  \jt{SIAM
  J. Appl. Math.}  \bvol{72}~(2),  \pg{689--711}.

\bibitem[Akylas(1984)]{akylas}
{\sc \au{Akylas, T.~R.}} \yr{1984}  \at{On the excitation of long nonlinear
  water waves by a moving pressure distribution}.  \jt{J.~Fluid Mech.}
  \bvol{141},  \pg{455--466}.

\bibitem[Alias(2014)]{alias}
{\sc \au{Alias, A.~B.}} \yr{2014} {\em Mathematical Modelling of Nonlinear
  Internal Waves in a Rotating Fluid\/}.  \publ{PhD Thesis}.

\bibitem[Berm{\'u}dez {\em et~al.\/}(2007)Berm{\'u}dez, Hervella-Nieto, Prieto
  \& Rodr{\i}]{bermudez2007optimal}
{\sc \au{Berm{\'u}dez, A.}, \au{Hervella-Nieto, L.}, \au{Prieto, A.} \&
  \au{Rodr{\i}, R.~\textit{et al.}}} \yr{2007}  \at{An optimal perfectly
  matched layer with unbounded absorbing function for time-harmonic acoustic
  scattering problems}.  \jt{J. Comp. Phys.}  \bvol{223}~(2),  \pg{469--488}.

\bibitem[Binder(2019)]{binderreview}
{\sc \au{Binder, B.~J.}} \yr{2019}  \at{Steady two-dimensional free-surface
  flow past disturbances in an open channel: Solutions of the {K}orteweg-de
  {V}ries equation and analysis of the weakly nonlinear phase space}.
  \jt{Fluids}  \bvol{4},  \pg{1--24}.

\bibitem[Binder {\em et~al.\/}(2013)Binder, Blyth \& McCue]{binder+mccue}
{\sc \au{Binder, B.~J.}, \au{Blyth, M.~G.} \& \au{McCue, S.~W.}} \yr{2013}
  \at{Free-surface flow past arbitrary topography and an inverse approach to
  wave-free solutions}.  \jt{IMA J.~ Appl. Math.}  \bvol{78},  \pg{685--696}.

\bibitem[Binder {\em et~al.\/}(2008)Binder, Dias \& Vanden-Broeck]{binder2008}
{\sc \au{Binder, B.~J.}, \au{Dias, F.} \& \au{Vanden-Broeck, J-M.}} \yr{2008}
  \at{Influence of rapid changes in a channel bottom on free-surface flows}.
  \jt{IMA J. Appl. Math.}  \bvol{73},  \pg{254--273}.

\bibitem[Binder {\em et~al.\/}(2005)Binder, Vanden-Broeck \&
  Dias]{binder+vanden}
{\sc \au{Binder, B.~J.}, \au{Vanden-Broeck, J-M.} \& \au{Dias, F.}} \yr{2005}
  \at{Forced solitary waves and fronts past submerged obstacles}.  \jt{CHAOS}
  \bvol{15},  \pg{037106 1--13}.

\bibitem[Boyd(2000)]{boyd1}
{\sc \au{Boyd, J.~P.}} \yr{2000} {\em Chebyshev and {F}ourier Spectral
  Methods\/}.  \publ{Dover Publications}.

\bibitem[Buttle {\em et~al.\/}(2018)Buttle, Pethiyagoda, Moroney \&
  McCue]{buttle2018three}
{\sc \au{Buttle, N.~R.}, \au{Pethiyagoda, R.}, \au{Moroney, T.~J.} \&
  \au{McCue, S.~W.}} \yr{2018}  \at{Three-dimensional free-surface flow over
  arbitrary bottom topography}.  \jt{J. Fluid Mech.}  \bvol{846},
  \pg{166--189}.

\bibitem[Camassa \& Wu(1991)]{camassa}
{\sc \au{Camassa, R.} \& \au{Wu, T. Y-T.}} \yr{1991}  \at{Stability of forced
  steady solitary waves}.  \jt{Phil. Trans. R. Soc. Lond.}  \bvol{10},
  \pg{429--466}.

\bibitem[Chardard {\em et~al.\/}(2011)Chardard, Dias, Nyguyen \&
  Vanden-Broeck]{chardard}
{\sc \au{Chardard, R.}, \au{Dias, F.}, \au{Nyguyen, H.~Y.} \&
  \au{Vanden-Broeck, J-M.}} \yr{2011}  \at{Stability of some stationary
  soutions to the forced {K}d{V} equation with one or two bumps}.  \jt{J.~Eng.
  Math.}  \bvol{70},  \pg{175--189}.

\bibitem[Choi \& Kim(2016)]{choi2016hyperbolic}
{\sc \au{Choi, H.} \& \au{Kim, H.}} \yr{2016}  \at{The hyperbolic relaxation
  systems for the forced {K}d{V} equations with hydraulic falls}.  \jt{Euro. J.
  Mech.-B/Fluids}  \bvol{58},  \pg{20--28}.

\bibitem[Craig \& Sulem(1993)]{craig1993numerical}
{\sc \au{Craig, W.} \& \au{Sulem, C.}} \yr{1993}  \at{Numerical simulation of
  gravity waves}.  \jt{J. Comp. Phys.}  \bvol{108}~(1),  \pg{73--83}.

\bibitem[Dias \& Vanden-Broeck(1989)]{dias1989open}
{\sc \au{Dias, F.} \& \au{Vanden-Broeck, J.-M.}} \yr{1989}  \at{Open channel
  flows with submerged obstructions}.  \jt{J. Fluid Mech.}  \bvol{206},
  \pg{155--170}.

\bibitem[Dias \& Vanden-Broeck(2002)]{dias+vanden-broeck}
{\sc \au{Dias, F.} \& \au{Vanden-Broeck, J.~M.}} \yr{2002}  \at{Generalised
  critical free-surface flows}.  \jt{J.~Eng. Math.}  \bvol{42},  \pg{291--302}.

\bibitem[Donahue \& Shen(2010)]{donahue}
{\sc \au{Donahue, A.~S.} \& \au{Shen, S. S-P.}} \yr{2010}  \at{Stability of
  hydraulic falls and sub-critical cnoidal waves in water flows over a bump}.
  \jt{J.~Eng Maths}  \bvol{10},  \pg{1007/s}.

\bibitem[Forbes(1988)]{forbes1988critical}
{\sc \au{Forbes, L.~K.}} \yr{1988}  \at{Critical free-surface flow over a
  semi-circular obstruction}.  \jt{J. Eng. Math.}  \bvol{22}~(1),  \pg{3--13}.

\bibitem[Forbes \& Schwartz(1982)]{forbes1982}
{\sc \au{Forbes, L.~K.} \& \au{Schwartz, L.~W.}} \yr{1982}  \at{Free-surface
  flow over a semicircular obstruction}.  \jt{J. Fluid Mech.}  \bvol{114},
  \pg{209--314}.

\bibitem[Forbes {\em et~al.\/}(2021)Forbes, Walters \&
  Hocking]{forbes2021ideal}
{\sc \au{Forbes, L.~K.}, \au{Walters, S.~J.} \& \au{Hocking, G.~C.}} \yr{2021}
  \at{Ideal planar fluid flow over a submerged obstacle: review and extension}.
   \jt{ANZIAM}  \bvol{63}~(4),  \pg{377--419}.

\bibitem[Grimshaw \& Maleewong(2016)]{grimshawsponge}
{\sc \au{Grimshaw, R. H.~J.} \& \au{Maleewong, M.}} \yr{2016}
  \at{Transcritical flow over two obstacles: Forced {K}orteweg--de {V}ries
  framework}.  \jt{J. Fluid Mech.}  \bvol{809},  \pg{918--940}.

\bibitem[Grimshaw \& Smyth(1986)]{grimshaw1986}
{\sc \au{Grimshaw, R. H.~J.} \& \au{Smyth, N.}} \yr{1986}  \at{Resonant flow of
  a stratified fluid over topography}.  \jt{J. Fluid. Mech.}  \bvol{169},
  \pg{429--464}.

\bibitem[Grimshaw {\em et~al.\/}(2007)Grimshaw, Zhang \& Chow]{grimshaw1}
{\sc \au{Grimshaw, R. H.~J.}, \au{Zhang, D-H.} \& \au{Chow, K.~W.}} \yr{2007}
  \at{Generation of solitary waves by transcritical flow over a step}.
  \jt{J.~Fluid Mech.}  \bvol{537},  \pg{235--254}.

\bibitem[Heil \& Hazel(2006)]{heil2006oomph}
{\sc \au{Heil, M.} \& \au{Hazel, A.~L.}} \yr{2006}  \at{oomph-lib--an
  object-oriented multi-physics finite-element library}.  \bt{In {\em
  Fluid-structure interaction\/}},  \pg{pp. 19--49}.  \publ{Springer}.

\bibitem[Heroux {\em et~al.\/}(2003)Heroux, Bartlett, Howle, Hoekstra, Hu,
  Kolda, Lehoucq, Long, Pawlowski, Phipps, Salinger, Thornquist, Tuminaro,
  Willenbring \& Williams]{herouxtrilnos}
{\sc \au{Heroux, M.}, \au{Bartlett, R.}, \au{Howle, V.}, \au{Hoekstra, R.},
  \au{Hu, J.}, \au{Kolda, T.}, \au{Lehoucq, R.}, \au{Long, K.}, \au{Pawlowski,
  R.}, \au{Phipps, E.}, \au{Salinger, A.}, \au{Thornquist, H.}, \au{Tuminaro,
  R.}, \au{Willenbring, J.} \& \au{Williams, A.}} \yr{2003}  \bt{An overview of
  trilinos}. {\em Tech. Rep.\/}.  \org{SAND2003-2927. Sandia National
  Laboratories}.

\bibitem[Holm {\em et~al.\/}(1985)Holm, Marsden, Ratiu \&
  Weinstein]{holm1985nonlinear}
{\sc \au{Holm, D.~D.}, \au{Marsden, J.~E.}, \au{Ratiu, T.} \& \au{Weinstein,
  A.}} \yr{1985}  \at{Nonlinear stability of fluid and plasma equilibria}.
  \jt{Phys. Rep.}  \bvol{123}~(1-2),  \pg{1--116}.

\bibitem[Keeler {\em et~al.\/}(2017)Keeler, Binder \&
  Blyth]{keelercriticalflow}
{\sc \au{Keeler, J.~S.}, \au{Binder, B.~J.} \& \au{Blyth, M.~G.}} \yr{2017}
  \at{On the critical free-surface flow over localised topography}.  \jt{J.
  Fluid Mech.}  \bvol{832},  \pg{73--96}.

\bibitem[Keeler {\em et~al.\/}(2021)Keeler, Blyth \& King]{keelernonlinear}
{\sc \au{Keeler, J.~S.}, \au{Blyth, M.~G.} \& \au{King, J.~R.}} \yr{2021}
  \at{Termination points and homoclinic glueing for a class of inhomogeneous
  nonlinear ordinary differential equations}.  \jt{Nonlinearity}  \bvol{34},
  \pg{532--561}.

\bibitem[Keeler {\em et~al.\/}(2019)Keeler, Thompson, Lemoult, Juel \&
  Hazel]{keeler2019invariant}
{\sc \au{Keeler, J.~S.}, \au{Thompson, A.~B.}, \au{Lemoult, G.}, \au{Juel, A.}
  \& \au{Hazel, A.~L.}} \yr{2019}  \at{The influence of invariant solutions on
  the transient behaviour of an air bubble in a hele-shaw channel}.  \jt{Proc.
  R. Soc. Lond. A}  \bvol{879},  \pg{1--27}.

\bibitem[Kuznetsov(1998)]{kuznetsov2013elements}
{\sc \au{Kuznetsov, Y.~A.}} \yr{1998} {\em Elements of Applied Bifurcation
  Theory (3rd Ed.)\/}.  \publ{Springer-Verlag}.

\bibitem[Li(2005)]{li2005overview}
{\sc \au{Li, X.~S.x}} \yr{2005}  \at{An overview of superlu: Algorithms,
  implementation, and user interface}.  \jt{ACM Transactions on Mathematical
  Software (TOMS)}  \bvol{31}~(3),  \pg{302--325}.

\bibitem[Page \& P{\u{a}}r{\u{a}}u(2014)]{page2014}
{\sc \au{Page, C.} \& \au{P{\u{a}}r{\u{a}}u, E.~I.}} \yr{2014}  \at{Time
  dependent hydraulic falls and trapped waves over submerged obstacles}.
  \jt{Phys. Fluids}  \bvol{26},  \pg{126604}.

\bibitem[Page \& Kerswell(2018)]{page2018koopman}
{\sc \au{Page, J.} \& \au{Kerswell, R.~R.}} \yr{2018}  \at{Koopman analysis of
  {B}urgers equation}.  \jt{Phys. Rev. Fluids}  \bvol{3}~(7),  \pg{071901}.

\bibitem[Page \& Kerswell(2019)]{page2019koopman}
{\sc \au{Page, J.} \& \au{Kerswell, R.~R.}} \yr{2019}  \at{Koopman mode
  expansions between simple invariant solutions}.  \jt{J. Fluid Mech.}
  \bvol{879},  \pg{1--27}.

\bibitem[Romate(1991)]{romate1991}
{\sc \au{Romate, J.~E.}} \yr{1991}  \at{Absorbing boundary conditions for free
  surface waves}.  \jt{J. Comp. Phys.}  \bvol{99},  \pg{135--145}.

\bibitem[Sandstede \& Scheel(2000)]{sandstede2000absolute}
{\sc \au{Sandstede, B.} \& \au{Scheel, A.}} \yr{2000}  \at{Absolute and
  convective instabilities of waves on unbounded and large bounded domains}.
  \jt{Physica D}  \bvol{145}~(3-4),  \pg{233--277}.

\bibitem[Tam {\em et~al.\/}(2015)Tam, Yu, Kelso \& Binder]{tam}
{\sc \au{Tam, A.~T.}, \au{Yu, Z.}, \au{Kelso, R.~M.} \& \au{Binder, B.~J.}}
  \yr{2015}  \at{Predicting channel bed topography in hydraulic falls}.
  \jt{Phys. Fluids}  \bvol{27},  \pg{112--106}.

\bibitem[Thompson {\em et~al.\/}(2014)Thompson, Juel \&
  Hazel]{thompson2013stability}
{\sc \au{Thompson, A.~B.}, \au{Juel, A.} \& \au{Hazel, A.~L.}} \yr{2014}
  \at{Multiple finger propagation modes in hele-shaw channels of variable
  depth}.  \jt{J. Fluid Mech.}  \bvol{746},  \pg{123--164}.

\bibitem[{\c{T}}ugulan {\em et~al.\/}(2022){\c{T}}ugulan, Trichtchenko \&
  P{\u{a}}r{\u{a}}u]{ctugulan2022three}
{\sc \au{{\c{T}}ugulan, C.}, \au{Trichtchenko, O.} \& \au{P{\u{a}}r{\u{a}}u,
  E.~I.}} \yr{2022}  \at{Three-dimensional waves under ice computed with novel
  preconditioning methods}.  \jt{J. Comp. Phys.}  \bvol{459},  \pg{111129}.

\bibitem[Wade {\em et~al.\/}(2014)Wade, Binder, Mattner \& Denier]{wade+binder}
{\sc \au{Wade, S.~L.}, \au{Binder, B.~J.}, \au{Mattner, T.~W.} \& \au{Denier,
  J.~P.}} \yr{2014}  \at{On the free surface flow of very steep forced solitary
  waves}.  \jt{J.~Fluid Mech.}  \bvol{739},  \pg{1--21}.

\bibitem[Wade {\em et~al.\/}(2017)Wade, Binder, Mattner \&
  Denier]{wade+binder2}
{\sc \au{Wade, S.~L.}, \au{Binder, B.~J.}, \au{Mattner, T.~W.} \& \au{Denier,
  J.~P.}} \yr{2017}  \at{Steep waves in free-surface flow past narrow
  topography}.  \jt{Phys. Fluids}  \bvol{29},  \pg{062107}.

\bibitem[Whitham(1974)]{whitham}
{\sc \au{Whitham, G.~B}} \yr{1974} {\em Linear and nonlinear waves\/}.
  \publ{John Wiley}.

\bibitem[Wu(1987)]{wu1987generation}
{\sc \au{Wu, T. Y-T.}} \yr{1987}  \at{Generation of upstream advancing solitons
  by moving disturbances}.  \jt{J. Fluid. Mech.}  \bvol{184},  \pg{75--99}.

\bibitem[Zakharov(1968)]{zakharov1968stability}
{\sc \au{Zakharov, V.~E.}} \yr{1968}  \at{Stability of periodic waves of finite
  amplitude on the surface of a deep fluid}.  \jt{J. Appl. Mech. Tech. Phys.}
  \bvol{9}~(2),  \pg{190--194}.

\end{thebibliography}

\end{document}